\documentclass{jpp}
\usepackage{graphicx}
\usepackage{multirow}
\usepackage{hyperref}

\usepackage[utf8]{inputenc}
\usepackage[T1]{fontenc}
\usepackage{amsmath}
\usepackage{amssymb}

\usepackage{nicematrix}
\usepackage{booktabs}
\usepackage{diagbox}
\usepackage[export]{adjustbox}

\newcommand{\ie}{{\it i.e.}}

\newcommand{\pelon}{\bar{e}}

\usepackage{unicode}
\newcommand{\csso}{\fontencoding{LECO}\selectfont\char215}
\newcommand{\iotaslash}{\hspace*{0.1em}\iota\hspace*{-0.45em}\text{\csso}}

\renewcommand{\t}{\hat{\pmb t}}
\newcommand{\n}{\hat{\pmb \kappa}}
\renewcommand{\b}{\hat{\pmb \tau}}

\shorttitle{A geometric approach to constructing quasi-isodynamic fields}
\shortauthor{G. G. Plunk and E Rodr\'{i}guez}

\title{A geometric approach to constructing quasi-isodynamic fields}

\author{G. G. Plunk\corresp{\email{gplunk@ipp.mpg.de}} and E. Rodr\'{i}guez}
\affiliation{Max Planck Institute for Plasma Physics, 17491 Greifswald, Germany}

\begin{document}

\maketitle

\begin{abstract}
The near-axis theory for quasi-isodynamic stellarator equilibria is reformulated in terms of geometric inputs, to allow greater control of the ``direct construction'' of quasi-isodynamic configurations, and to facilitate understanding of the space of such equilibria.  This  includes a method to construct suitable magnetic axis curves by solving Frenet-Serret equations, and an approach to controlling magnetic surface shaping at first order (plasma elongation), which previously has required careful parameter selection or additional optimization steps.  The approach is suitable for studying different classes of quasi-isodynamic stellarators including different axis ``helicities'' and topologies (\textit{e.g.} knotted solutions), and as the basis for future systematic surveys using higher order near-axis theory. As an example application, we explore a family of configurations with per-field-period axis helicity equal to one half, demonstrating an approximate scaling symmetry relating different field period numbers.
\end{abstract}

\section{Introduction}

The `indirect' or `inverse-coordinate' approach to solving the magnetic equilibrium problem near the magnetic axis of a stellarator is formulated in magnetic coordinates to allow important symmetries of the magnetic field (quasi-symmetry or omnigenity) to be easily enforced \citep{garren-boozer-1}.  Compared with `direct' approach of Mercier \citep{mercier-near-axis,Solovev1970} (where magnetic coordinates are an output of the calculation), the indirect approach is in some ways less intuitive, as many of the inputs are given in abstract quantities that do not have such a simple geometric interpretation.  For the case of quasi-isodynamic (QI) magnetic equilibria \citep{gori-lotz-nuehrenberg}, the issue is more severe, as the quality of solutions is sensitive to basic geometric features, {\em e.g.} the axis shape and surface elongation, that have proved difficult to control.  This can be overcome, to some extent, by careful choice of input parameters \citep{camacho-mata_plunk_2022, Camacho-Mata_Plunk_2023}, or by an additional optimization step, within the space of near-axis solutions \citep{Jorge2022}.

An alternative approach, as pursued here, is to choose the inputs of the theory to be more geometrical, starting with the magnetic axis itself (zeroth order), solved using Frenet-Serret equations \citep{frenet1852courbes, animov2001differential}, and proceeding to the elliptical shaping parameters of the magnetic surfaces (first order).  This approach can be viewed as an attempt to include some of the benefits of Mercier's approach to make the inverse formulation more geometrically transparent.

The method was presented by \cite{plunk-simons-2023} but so far not described in detail elsewhere.  Several papers have been written that use, in one form or another, configurations constructed using this methodology \citep{Goodman_PRXEnergy_2024, Rodriguez_Helander_Goodman_2024, Rodriguez_Plunk_Residual_2024, Rodríguez_Plunk_Jorge_2025, hindenlang_plunk_maj_Gframe_2024, Plunk_figure-8_2025}, giving further motivation to publish details about it.

In this work, we describe this geometric method to construct QI fields, demonstrate its flexibility to explore different classes of QI, characterized by magnetic axis topology, and orders of axis curve flattening at field extrema.  We begin with some essential background (Section \ref{sec:nae_theory}), proceed to details of the method (Section \ref{sec:method}) and then describe the overall `construction' of the full solution to first order in the expansion, with examples.  Finally, a survey of the conventional class of QI (``half-helicity'') is performed, with emphasis on geometric properties and scaling behavior with field period number $N$.  In the appendices we describe the connection between our `hybrid' indirect/direct approach and the original work of Mercier.

\section{Zeroth order near axis theory: Magnetic axis shape and on-axis field strength} \label{sec:nae_theory}

The near-axis expansion (NAE) \cite{plunk_landreman_helander_2019, camacho-mata_plunk_2022, Jorge2022,Rodriguez_Plunk_2023_Higher_QI,Rodríguez_Plunk_Jorge_2025} provides a theoretical framework for finding approximate stellarator equilibrium fields, valid in the neighborhood of the magnetic axis.  Global solutions with finite radius can then be ``directly constructed'' \cite{landreman-sengupta, plunk_landreman_helander_2019} by evaluating the asymptotic description near the magnetic axis, at a specified distance (average minor radius).  The resulting plasma boundary shape can be provided as input to a magnetohydrodynamic (MHD) equilibrium code like VMEC \citep{Hirshman}, GVEC \citep{gvec-2019} or DESC \citep{dudt_DESC_2020}, for further investigation.  In this paper we perform the expansion up to first order, starting in this section with the zeroth-order part, the magnetic axis and on-axis magnetic field strength.

We will focus on the simple case of a single magnetic trapping well per field period, {\em i.e.} with a single minimum in the field strength on axis, $B_0(\varphi)$.  We also assume stellarator symmetry \citep{dewar1998stellarator}, implying that the field extrema coincide with stellarator-symmetric points.  By the convention used here, these are located at $\varphi_{\mathrm{max}}^{(i)} = 2\pi i/N$ for maxima and $(2 i + 1)\pi/N$ for minima, where $i$ is an integer and $\varphi$ is the Boozer toroidal angle \citep{boozer1981plasma,dhaeseleer}.

The magnetic axis is a closed space curve representing the central magnetic field line of the stellarator. The geometry of this curve affects the behavior of the field. Most importantly, the curvature of the axis is directly linked to the gradients of the field strength.  For this reason, the axis curves of a quasi-isodynamic stellarator have "flattening points" (points of zero curvature) at extrema of the magnetic field strength along the axis; this flattening is necessary for poloidal closure of the contours of constant field strength $|\mathbf{B}|$ \citep{plunk_landreman_helander_2019,Rodríguez_Plunk_Jorge_2025}.
\par
This constraint on the curvature already makes the problem of constructing suitable curves a non-trivial part of constructing a QI configuration. In previous works on near-axis QI stellarators, magnetic axis curves that satisfy these flattening conditions were found by applying linear constraints to Fourier coefficients of the curve in cylindrical coordinates (see \cite{camacho-mata_plunk_2022} and Appendix~\ref{sec:app_torsion} here).  Curves constructed this way are closed and smooth by construction, and satisfy the required flattening conditions.  The presence of these straight sections, however, make such curves sensitive to deformations.  The torsion in particular is prone to singularities near flattening points (see Appendix~\ref{sec:app_torsion} for a detailed discussion).  Because torsion and other geometric properties of the axis curve directly control magnetic field properties such as the rotational transform and flux surface shaping, a level of control on these properties is desired that the conventional approach just described does not easily provide (see some further discussion in Appendix~\ref{sec:app_torsion}). 
\par
In this work, to address the issues just discussed, we propose a method to construct the axis curve by prescribing curvature, $\kappa$, and torsion, $\tau$, as functions of the arc length $\ell$ (for simplicity $\ell\in[0,2\pi)$) along the curve, and solve the Frenet-Serret (FS) equations \citep{frenet1852courbes, animov2001differential},
\begin{equation}
    \frac{d\mathbf{x}_0}{d\ell}=\t,\quad
    \frac{d\t}{d\ell}=\kappa\n,\quad
    \frac{d\n}{d\ell}=-\kappa\t + \tau \b,\quad
    \frac{d\b}{d\ell}=-\tau\n,  \label{eqn:FS_eqns}       
\end{equation}
where ${\bf x}_0$ is the axis curve, $\t$ is its tangent, and $(\n,\b)$ complete the orthonormal frame. For any functions $\kappa$ and $\tau$, Eqns.~\ref{eqn:FS_eqns} may be integrated to obtain a curve, which is unique up to translations and rotations, {\em i.e.} initial conditions for the curve $\mathbf{x}_0$ and its frame \citep[Thrm.~3.4.1]{banchoff2022differential}\citep[Ch.~1-5]{do2016differential}. 

Treating curvature as an input to the curve construction ensures that the curve can have the desired flattening points. We can categorize different curves in terms of the order of the zeros of curvature, which we denote $(i,j)$, referring to the field maxima and minima respectively. Across some of these points, the curve undergoes {\em inflection}, by which we mean that the vector $d\t/d\ell$ flips direction discontinuously. In a QI field, inflection is mandatory at locations of minimum field strength \citep{plunk_landreman_helander_2019, Rodriguez_Plunk_2023_Higher_QI}. For this reason $\kappa$ is defined as ``signed curvature'', which crosses zero at such points.  By stellarator symmetry the signed curvature is therefore odd about locations of field minima, while torsion is an even function. Furthermore $(\t, \n, \b)$ must be defined as the \textit{signed} Frenet frame \citep{plunk_landreman_helander_2019,Rodríguez_Plunk_Jorge_2025,carroll2013-journal}, {\em i.e.} $(\t, \n, \b) = (\t^{F}, s_\kappa \n^{F}, s_\kappa \b^{F})$ where $s_\kappa = \mathrm{sign}(\kappa)$ and $(\t^{F}, \n^{F}, \b^{F})$ is the conventionally defined Frenet frame, {\em i.e.} $\t^{F} = \mathbf{x}_0^\prime$, $\n^{F}=\mathbf{x}_0^{\prime\prime}/|\mathbf{x}_0^{\prime\prime}|$ and $\b^{F} = \t^{F}\times \n^{F}$, defined in the limiting sense approaching those points where  $|\mathbf{x}_0^{\prime\prime}| = 0$ and where primes denote derivatives with respect to $\ell$.  

The signed Frenet frame is by our convention continuous within a field period, {\em i.e.} $\ell \in [2\pi n/N,2\pi(n+1)/N)$, where $n$ is an integer. In particular, field periods are taken to begin and end at locations of maximum field strength, and the FS equations are solved within a single field period with the full curve determined by continuation, using the $N$-fold symmetry.  

An important property of the magnetic axis curve is its \textit{helicity}, {\em i.e.} the number of times the signed normal rotates about the axis \citep{landreman2019,rodriguez2022phases,mata2023helicity,Rodríguez_Plunk_Jorge_2025} (also self-linking number \cite{moffatt1992helicity,fuller1999geometric,fuller-writhe}). We define the per-field-period helicity as $m = M/N$, where $M$ the total number of turns for all field periods. The per-period helicity can be any integer multiple of $1/2$ \citep{Camacho-Mata_Plunk_2023,Rodríguez_Plunk_Jorge_2025}. There is no closed-form expression for helicity in terms of curvature and torsion, so the curve must first be constructed (by solving the FS equations) to compute the helicity.\footnote{If analyticity is assumed at flattening points, then it can be shown that $i+j$ must be odd for the helicity to be a half integer, {\em i.e.} $m = n + 1/2$ with $n$ an integer \citep{Camacho-Mata_Plunk_2023,Rodríguez_Plunk_Jorge_2025}.}

Although the FS equations can, for arbitrary choice of $\tau$ and $\kappa$, be integrated to obtain a space curve, this curve will generally not be closed.  A further optimization step is therefore required to achieve that \citep{garren-boozer-1}. The remainder of this section is dedicated to discussing the details of how to achieve axis closure.  First, to ensure that axis curves can be constructed uniquely (excluding differences due to trivial translational and rotational freedoms), we will adopt a number of conventions:\newline

\begin{enumerate}
    \item Field periods are taken to begin and end at locations $\phi = 2 \pi n/N$, where $\phi$ is the cylindrical angle, and $n$ takes consecutive integer values.  Lines of stellarator symmetry lie at these locations, as well as at their midpoints, $\phi = \pi n/N$.\label{curve-conventions-1} 
    \item The axis curve is centered at $x=y=z=0$; for $N=1$ the curve is positioned such that this point is halfway between the locations of the field maximum and minimum.\label{curve-conventions-2}
    \item The axis curve begins along the $x$-axis (the $y$ and $z$ components of ${\mathbf x}_0$ are zero at $\ell = 0$), {\em i.e.} ${\mathbf x}_0|_{\ell = 0} = (x_0, 0 , 0)$.\label{curve-conventions-3}
    \item For the case $N = 1$ an additional rotational freedom (about a single axis of stellarator symmetry) is fixed by taking ${\bf x}_0\cdot \hat{\bf z} = 0$ at $\phi = \pi/2$. This extra constraint is needed as the number of symmetry points (\ref{curve-conventions-1}), only two, are insufficient to uniquely define a plane.\label{curve-conventions-4}
\end{enumerate}

\par 

\subsection{Closure criteria for $N = 1$}

The method to numerically obtain closed space curves is to tune the curvature and torsion until a number of ``closure criteria'' are satisfied to sufficient accuracy \citep{garren-boozer-1}.   Closure includes the three scalar criteria corresponding to ${\bf x}_0|_{\ell = 0} = {\bf x}_0|_{\ell = 2\pi}$ (the curve being closed), but also criteria associated with the periodicity of the frame (the frame being aligned), $\t|_{\ell = 0} = \t|_{\ell = 2\pi}$, {\it etc}. The ``endpoint'' quantities are computed by integrating the FS equations, Eqs.~(\ref{eqn:FS_eqns}). The case of fractional helicity ({\it i.e.} $m$ an integer plus $1/2$) is a special one, as the periodicity conditions that the frame satisfies can involve a sign flip.  For the $N=1$ case the signed normal $\n$ (and binormal, $\b$) will satisfy $\n|_{\ell = 0} = -\n|_{\ell = 2\pi}$ if the helicity is fractional in this sense.  

There is a certain amount of redundant information in the closure constraint of the frame, which is by construction orthonormal. In particular, the minimal number of scalar criteria necessary to guarantee alignment of the frame is much lower than the 9 scalar components of the frame vectors.  In fact it should only be three conditions, corresponding to the number of Euler angles necessary to fully describe the orientation of a rigid body.  Thus the closure for the most general class of $N=1$ curves, would seem to require six free ``shaping'' parameters in the functions $\kappa$ and $\tau$ (as many as closure conditions).  However, to simplify things a bit, we only consider stellarator symmetric curves here.

To help reduce the number of necessary closure conditions we note that stellarator symmetry constrains initial values for the frame and position of the curve, following convention (\ref{curve-conventions-3}). Consider a geometric consequence of the helicity of the axis: At the minima of the magnetic field, assuming stellarator symmetry, the inflection implies $\b$ is parallel to $\hat{\bf R}$ ($\phi = \pi/N$, {\it etc.}) \citep{camacho-mata_plunk_2022,Rodríguez_Plunk_Jorge_2025} -- otherwise, the curvature would cause the axis to move radially outward on one side of the minimum and radially inward on the other side, violating stellarator symmetry. Hence, for integer helicity $m$, $\b$ must also be parallel to $\hat{\bf R}$ at $\ell=0$ (the frame will make an integer number of half turns between field maxima and minima), while for fractional helicities, $\n$ is parallel to $\hat{\bf R}$ at such locations (a relative quarter turn will remain). So the choice of alignment of the frame when initialising the Frenet construction has important influence on the helicity of the curve. In short, $\n$ or $\b$ can be aligned with $\hat{\mathbf{x}}$ at $\ell=0$ depending on the desired helicity of the solution.  In addition to the normal components of the Frenet-Serret frame, there is also rotational freedom in the tangent about $\hat{\mathbf{x}}$ at $\ell=0$. To uniquely choose this freedom, we shall rotate the final curve so as to satisfy condition \ref{curve-conventions-4}.

With these initial conditions on the frame, we can then evaluate the closure criteria, which, using stellarator symmetry, can be applied at $\ell = \pi$.  From the symmetry of Eq.~(\ref{eqn:R_and_Z}), the $\hat{\bf x}$ component of ${\bf x}_0$ is invariant under $\ell\rightarrow\pi-\ell$, and closure of the curve only requires the other (odd) components to be zero ({\em i.e.} requires the point at $\ell=\pi$ to lie along the x axis):
\begin{subequations}
\begin{eqnarray}
    \hat{\bf y}\cdot {\bf x}_0|_{\ell = \pi} = 0,\\
    \hat{\bf z}\cdot {\bf x}_0|_{\ell = \pi} = 0. \label{eq:N1_z.x0}
\end{eqnarray}
Similar arguments may be extended to the tangent and normal components of the frame (using Eq.~(\ref{eqn:aux_cross_r}) instead of Eq.~(\ref{eqn:R_and_Z}) for the symmetry arguments). In both cases, the vectors must live in the plane normal to $\hat{\mathbf{R}}$ (or the x-axis by \ref{curve-conventions-3}), so that the last two closure conditions are this way obtained
\begin{eqnarray}
    \hat{\bf x}\cdot \t|_{\ell = \pi} = 0,\\
    \hat{\bf x}\cdot \n|_{\ell = \pi} = 0.
\end{eqnarray}
\end{subequations}
With four conditions to satisfy, we allow for four free ``shaping'' parameters for the magnetic axis curvature and torsion.  An example of a suitable parametrization is as follows

\begin{align}
    \kappa =&  \left( \kappa_1 \sin(N \ell) + \kappa_2 \sin(2 N \ell) \right)\cos^2\left(\frac{N \ell}{2}\right) \sin\left(\frac{N \ell}{2}\right),\label{eq:kappa-N1}\\
    \tau = & \tau_0 + \tau_1 \cos( N \ell) + \tau_2 \cos(2 N \ell),\label{eq:tau-N1}
\end{align}
This form of $\kappa$ corresponds to curves of the class $(2,3)$: $2$nd order zeros at the field maxima and $3$rd order zeroes at the minima.  

In practice, of the five parameters, four can be used to satisfy closure criteria, leaving one left to define a one-parameter family of curves. We iteratively integrate FS equations, using a root solver, to obtain suitable coefficients.  The desired helicity class (integer or half-integer) fixes the initial condition for the frame, but the actual helicity is only determined by inspection of the final solution.  This method is sensitive to the initial guess (for shaping parameters) and not guaranteed to find all desired solutions even within a particular class, {\it i.e.} given field period number $N$ and helicity $m$.  

The curve obtained by this procedure will generally not comply with conventions \ref{curve-conventions-2} and \ref{curve-conventions-4}, and thus a final translation along the $x$-axis and rotation about the $x$-axis are applied.

\subsection{Closure criteria for $N \geq 2$}

The case $N \geq 2$ yields significant simplification as compared to the case $N = 1$.  We integrate the FS equations over a single field period $\ell \in [0, 2\pi/N]$, and derive a set of conditions to apply at the ends of this domain, with these conditions designed to ensure both that the curve itself closes (after $N$ field periods), and that the frame is suitably periodic.

Let us describe how to use field-periodicity to reduce the number of closure criteria.  First we start with a generally open curve, obtained by integrating Eqns.~\ref{eqn:FS_eqns}, with its initial position along the x-axis following convention (\ref{curve-conventions-3}).  Rotational freedom about the $x$-axis can be used to ensure the end of this curve lies in the $x$-$y$ plane (which is required by periodicity), {\em i.e.} defining ${\bf x}_0|_{\ell = 2\pi/N} = (x_1, y_1, z_1)$, we have $z_1 = 0$.   Then, rotational freedom about an axis parallel to the z-axis, running through the initial point $(x_0, 0, 0)$, and translational freedom ($x_0$) can be used to ensure that the end of the curve is located at $\phi = 2\pi/N$, {\em i.e.} $y_1/x_1 = \tan(2\pi/N)$.  Therefore the closure of the curve ${\bf x}_0|_{\ell = 2\pi} = {\bf x}_0|_{\ell = 0}$ is achieved simply by $N$-fold periodic extension, {\it i.e.} by connecting $N$ copies of the single-period curve end-to-end.  With closure satisfied, only the continuity of the frame is left to be enforced.

\begin{figure}
    \centering
    \includegraphics[width=0.4\linewidth]{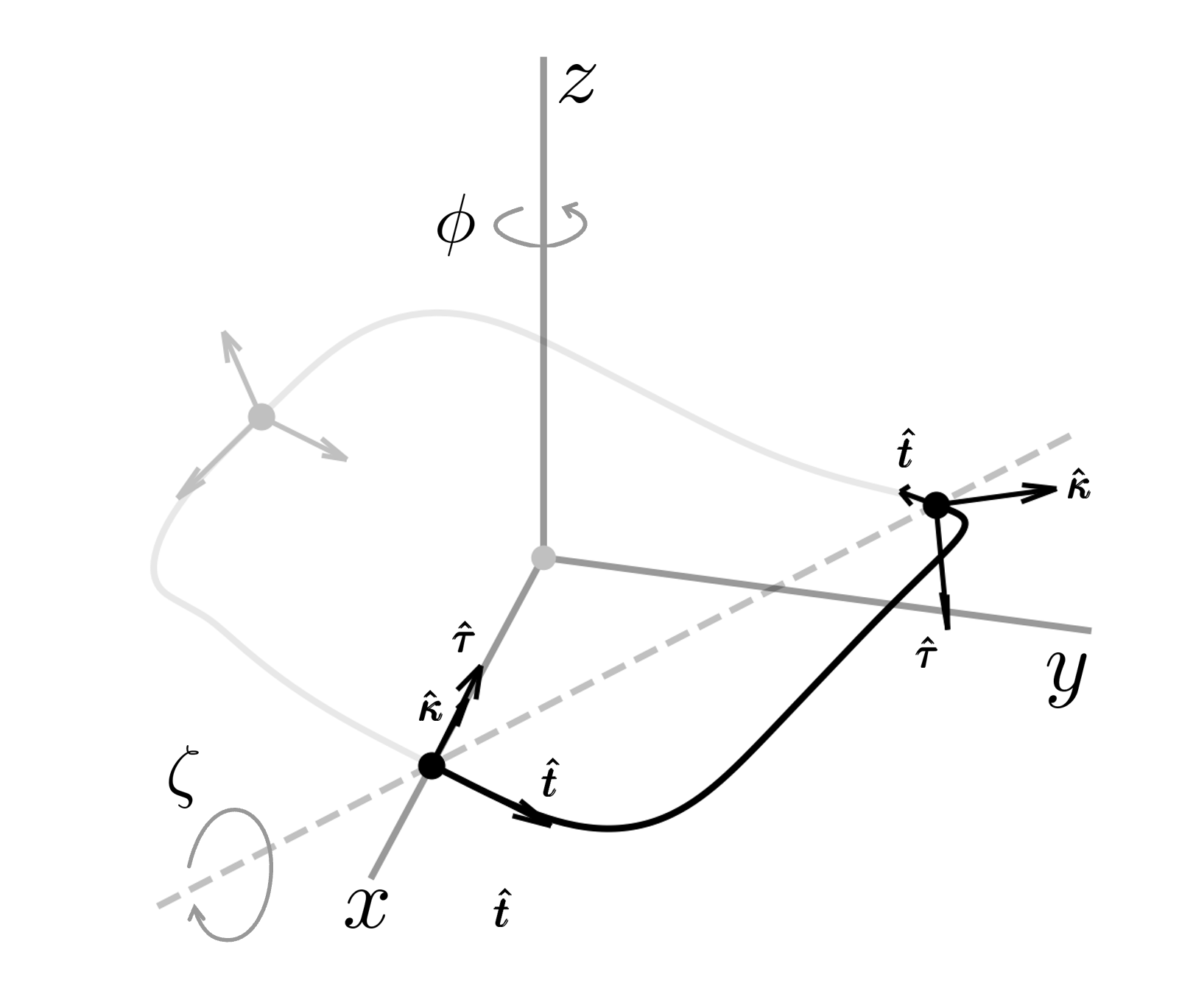}
    \caption{\textbf{Diagram illustrating elements of curve closure.} Example of a $N=3$ half-helicity curve with the main geometric elements considered for the closure of the axis, including the convention. In black, one field period of the axis from $\ell=0$ to $\ell=2\pi/3$, continued in gray. The $x,~y,~z$ frame is given as a reference.}
    \label{fig:closure_axis}
\end{figure}

At this stage (see Figure~\ref{fig:closure_axis}), there remains a single rotational freedom of the axis segment that preserves the location of the initial and final points (${\bf x}_0|_{\ell = 0}$ and ${\bf x}_0|_{\ell = 2\pi/N}$), namely rotation about the line connecting these two points, by an angle $\zeta$.
Because this rotation is not aligned with the $x$-axis, this rotation will actually change the initial orientation of the frame, and thus we can use this degree of freedom towards alignment of the frame.  To completely align the frames at the end and beginning of the period (up to a sign to allow for half helicity curves), we impose three scalar constraints (\ie, three Euler angles of the frame):
\begin{subequations}
\begin{eqnarray}
    {\n|_{\ell=2\pi/N}\cdot {\mathcal R}_{z}(2\pi/N)\cdot\t|_{\ell=0}} = 0,\\
     {\b|_{\ell=2\pi/N}\cdot {\mathcal R}_{z}(2\pi/N)\cdot\t|_{\ell=0}} = 0,\\
      {\b|_{\ell=2\pi/N}\cdot {\mathcal R}_{z}(2\pi/N)\cdot\n|_{\ell=0}} = 0,
\end{eqnarray}
\end{subequations}
where ${\mathcal R}_{z}(2\pi/N)$ is the operator that performs rotation by the angle $2\pi/N$ about the $z$-axis so that the frame at the endpoint can be compared to the initial one. 

To satisfy the three constraints, $\zeta$ provides one degree of freedom so two further free shaping parameters (by simple counting argument) are required as part of the curvature and torsion functions $\kappa$ and $\tau$ (consistent with Appendix~B of \cite{garren-boozer-2}); as an example, we consider the following forms
\begin{align}
    \kappa = & \kappa_1 \cos^2\left(\frac{N \ell}{2}\right) \sin\left(\frac{N \ell}{2}\right)\sin(N \ell),\label{eq:kappa-N2}\\
    \tau = & \tau_0 + \tau_1 \cos( N \ell),\label{eq:tau-N2}
\end{align}
where $\kappa_1$, $\tau_0$ and $\tau_1$ are constants.  This form of $\kappa$ (used for the figure-8 configuration of \cite{Plunk_figure-8_2025}) is also of the $(2,3)$ class. Two of the three free constants are determined by satisfaction of the closure conditions, leaving a single parameter to parametrize a family of curves. Just like the $N=1$ case, the method used for $N \geq 2$, and in particular the search for appropriate parameters, is sensitive to the initial guess, but we have not yet found evidence of multiple branches of solutions within a single class (given field period number $N$ and helicity $m$).

As a final consideration we note that, although we are interested in the case of stellarator symmetry, the above methodology for curve construction is actually capable of treating the general problem where symmetry is broken, by simply using more general (non-symmetric) curvature and torsion functions.  This is because stellarator symmetry actually requires the initial frame to have a particular alignment with respect to $\hat{\mathbf{x}}$, so that value of $\zeta$ obtained by the solution method is automatically constrained to satisfy this given symmetric curvature and torsion functions.  For non-symmetric curvature and torsion, $\zeta$ instead provides a continuous freedom for parameterizing the bigger space of non-symmetric curves.

\subsection{Complete zeroth order solution}

Once the an axis curve has been specified, the solution to zeroth order in the NAE is completed by also specifying the magnetic field strength dependence along the magnetic axis.  Stellarator symmetry requires an even function, for instance

\begin{equation}
  B_0 = 1 + \Delta \cos(N \ell) + \Delta \cos(2N \ell)/4.\label{eq:B0}
\end{equation}
defined such that $B_0^{\prime\prime} = 0$ at field strength minima, {\em e.g.} $\ell=\pi/N$, for reasons of stability and maximum-$\mathcal{J}$ \citep{Rodriguez_Helander_Goodman_2024, Plunk_figure-8_2025}.  The parameter $\Delta = (B_0(0)-B_0(\pi/N))/2$ controls the mirror ratio; note the mean $\langle B_0 \rangle_\ell = 1$.  Finally, the relationship between axis length and Boozer toroidal angle $\varphi$ is determined \citep{landreman-sengupta,plunk_landreman_helander_2019}:
\begin{equation}
\frac{d\varphi}{d\ell}= \frac{B_0}{|G_0|},
\hspace{0.3in}
|G_0| = \frac{N}{2\pi}\int_0^{2\pi/N} B_0\,d\ell, \label{eq:l-varphi}
\end{equation}
where $G_0$ is the zeroth-order Boozer poloidal current function.

\section{First order: Magnetic surface shaping}\label{sec:first-order}

To complete the construction of the configuration we must solve the so-called `$\sigma$-equation' \citep{garren-boozer-1, landreman-sengupta-plunk}, depending on several given inputs,

\begin{equation}
    \sigma^{\prime} + (\iotaslash_0 -\alpha^{\prime})\left(\sigma^{2}+ 1 + \bar{e}^2 \right) + 2 G_{0} \tau \bar{e}/B_0 = 0,\label{eq:sigma}
\end{equation}
where $\alpha$ and $\bar{e}$ are functions of $\varphi$ \citep{plunk_landreman_helander_2019,camacho-mata_plunk_2022,Rodríguez_Plunk_Jorge_2025}, and $\iotaslash_0$ is the rotational transform on axis; note we have assumed zero toroidal current for simplicity.  The solution to this equation determines the surface shape via the components of the first order coordinate mapping,

\begin{align}
    X_{1} &= \sqrt{\frac{2\bar{e}}{B_0}} \cos{[\theta - \alpha (\varphi) ]}, \label{eqn:X1} \\
    Y_{1} &= \sqrt{\frac{2}{B_0\bar{e}}}  \left( \sin{[\theta - \alpha (\varphi)]} + \sigma(\varphi) \cos{[\theta - \alpha (\varphi) ]}   \right), \label{eq:Y1}
\end{align}
which enters the coordinate mapping as follows

\begin{equation}
    \mathbf{x} \approx \mathbf{x}_{0} + \epsilon \left( X_{1}\mathbf{n} + Y_{1}\mathbf{b} \right),
\end{equation}
where $\epsilon=\sqrt{\psi}$.\footnote{This pseudo-radial coordinate differs by a factor of $\sqrt{2/\bar{B}}$ compared to some others \cite{landreman-sengupta,Rodríguez_Plunk_Jorge_2025}.} The function $\sigma$ controls rotation of the cross-sections, as well as the elongation of the surrounding elliptical cross-sections \citep{rodriguez2023magnetohydrodynamic}. Elongation (of the elliptical cross sections) is a particularly sensitive feature of QI configurations that are derived from NAE theory. The difficulty arises in maintaining the poloidal closure of the contours of $|\mathbf{B}|$, which can most easily be achieved by simply squeezing the ellipses along the normal direction to the axis (reducing poloidal variation of the field). Formally, the lack of control of elongation comes from $\sigma$ being an unknown of Eq.~(\ref{eq:sigma}) (alongside the rotational transform), and thus elongation, which depends on $\sigma$, has so far been obtained as an output of the construction. Although the control of torsion has been shown to be a plausible way of limiting this shaping \citep{camacho-mata_plunk_2022,Camacho-Mata_Plunk_2023} by limiting the size of $\sigma$, we here consider an alternative approach.

We propose a way to solve the $\sigma$ equation using elongation as input, instead of the function $\bar{e}$, which is used conventionally \cite{landreman-sengupta-plunk, plunk_landreman_helander_2019}. The approach can be regarded as a hybrid with Mercier's direct approach, in which the shape of flux surfaces is directly controlled in the near-axis description (see some further discussion of this in Appendix~\ref{sec:app_merc}).  The elongation $E(\varphi)$ of the elliptical cross section is

\begin{align}
    E(\varphi) &= \frac{1}{2}\left(\rho + \sqrt{\rho^2-4}\right),\label{eq:E-eqn}\\
    \quad \rho &= \bar{e} + \frac{1}{\bar{e}}(1 + \sigma^2).\label{eq:rho-eqn}
\end{align}
The ``elongation profile'' (see a geometric interpretation in Appendix~\ref{sec:app_merc}) is thus set by the function $\rho$, which for the class of configurations considered here is chosen to have the following form

\begin{equation}
\rho = \sum_n \rho_n \cos(n N \varphi),\label{eq:elongation-input-form}
\end{equation}
where $\rho_n$ are constants chosen such that $\rho \geq 2$ is satisfied for all $\varphi$. 

There is one more function that is part of the input at first order, namely the function $\alpha(\varphi)$, see Eqs.~(\ref{eqn:X1})-(\ref{eq:Y1}) and its appearance in Eq.~(\ref{eq:sigma}).  This phase angle controls the deviation of the field from the ideal QI limit 
\citep{plunk_landreman_helander_2019}.  However, to minimize the deviation from QI, there is little freedom in this function when stellarator symmetry is assumed, and it has so far seemed adequate to assume a sufficiently smooth form that avoids twist in the configurations \citep{camacho-mata_plunk_2022}.  A sufficiently high degree of smoothness also ensures good behavior at higher order NAE, as described in \cite{Rodríguez_Plunk_Jorge_2025}.

Because our method of solving the $\sigma$ equation requires both variation in $\bar{e}$ and the ability to enforce QI symmetry by choice of $\rho$, it does not seem applicable to the case of quasi-symmetry; indeed it is apparently limited only to the case of stellarator symmetric QI stellarators.

\subsection{Solution method}\label{sec:method}

As is generally the case with the inverse approach to near-axis theory, the first order problem may be solved by finding values of $\iota_0$ for which the solution of Eqn.~\ref{eq:sigma} yields periodic solutions of $\sigma$, {\em i.e.} $\sigma(0) = \sigma(2\pi/N) = 0$, assuming stellarator symmetry.  The key difference here is the use of $\rho$ instead of $\bar{e}$ as an input.  We do not attempt here to show that this alternative problem specification is well-posed, as it actually is not, with some inputs yielding no solutions.  However, the solution method is well-behaved in the sense that solutions can be rapidly found in a similar way as conventionally done (using $\bar{e}$ as an input), and the boundary in the input parameter space where solutions become invalid is identifiable, as described below.

To obtain the desired form of Eqn.~\ref{eq:sigma} (see Eq.~(\ref{eqn:sigma_rho})), we simply need to express $\bar{e}$ in terms of $\sigma$ and $\rho$, 
\begin{eqnarray}
    \bar{e} = \frac{1}{2}\left(\rho - \sqrt{\rho^2 - 4(1+\sigma^2)} \right),\label{eq:ebar-vs-rho}
\end{eqnarray}
and substitute it into Eqn.~\ref{eq:sigma} to eliminate $\bar{e}$.  Here we find two limitations of the approach. First, we have chosen the smaller root for $\bar{e}$, corresponding to elongation of the ellipses in the {\it conventional} sense, {\it i.e.} in the binormal direction. This is a limitation of the method, excluding cases where elongation passes from the conventional to unconventional (elongated in the direction of the normal vector) sense. This does not appear to be a practical limitation, as the overwhelming majority of cases studied so far have conventional elongation that only ever approaches $1$ at specific locations in $\varphi$.

A more significant downside of the method is that, even with $\rho \geq 2$, a real solution to the equation may not exist. This occurs whenever $\sigma$ is sufficiently large, which by inspecting Eqn.~\ref{eq:sigma} will tend to happen when the final source term is large, {\it i.e.} when the axis torsion is large, or when $\bar{e}$ is large.  However, reducing $\rho$, though it reduces $\bar{e}$, exacerbates the problem in Eqn.~\ref{eq:ebar-vs-rho}.  In practice this issue limits how small $\rho$ (and therefore plasma elongation) can be taken to obtain valid solutions for $\sigma$.  To cope with this, the requirement 
\begin{equation}
\rho^2 \geq 4 (1+\sigma^2),\label{eq:rho-crit}
\end{equation}
which depends on $\sigma$, must be enforced by checking numerical solutions.  Overall we have lost the uniqueness and existence properties of the standard approach to the solution of the $\sigma$-equation \citep{landreman-sengupta-plunk}, but have gained control of the geometric shaping.

We remark that the use of elongation explicitly in the solution process is reminiscent of the original near-axis expansion of \cite{mercier-near-axis}.  We have investigated this connection in some depth but leave these details for Appendix~\ref{sec:app_merc}).

\section{Constructing QI configurations}

We now turn to practical considerations for the actual construction of full solutions to first order.  The inputs at first order are considered as 
\vspace{0.5\baselineskip}
\begin{enumerate}
    \item the shape of the magnetic axis ${\bf x}_0$ (provided in terms of $\kappa$ and $\tau$),
    \item the form of the on-axis magnetic field strength $B_0(\varphi)$, and
    \item the elongation profile $\rho(\varphi)$.
\end{enumerate}
\vspace{0.5\baselineskip}
\noindent During construction of the magnetic axis, we also fix key quantities like the field periodicity $N$, the orders of zeros of the signed curvature at field maxima and minima, and axis helicity $m$.\footnote{Helicity is not an input in the sense that for instance $\rho(\varphi)$ is.   It can be controlled by restricting shaping parameters to a region in the neighborhood of known solutions with desired helicity, and checking the helicity after the FS problem is solved.}  For this paper, we focus on the forms of $\kappa$, $\tau$ and $B_0$ given in Section \ref{sec:nae_theory}, meaning that the orders of curvature zeros are $2$ and $3$ at the maxima and minima respectively. Many other forms are possible (and many have been implemented) but we will not attempt to list them here.  We note that an even order at field maxima ($2$) is consistent with analyticity only for the case of fractional helicity ($m = 1/2, 3/2, ...$) \citep{camacho-mata_plunk_2022,Rodríguez_Plunk_Jorge_2025}.  For present purposes analyticity is not a major concern, and it is noteworthy that the method is capable of treating more general (non-analytic) classes of curves.  It is even possible to have even orders of zero at inflection points as a sign change in the curvature is sufficient for inflection of the curves obtained from solving the FS Eqns.

To illustrate the flexibility of our method, we show a number of configurations in Table \ref{tab:Nm-Grid} representing a range of values of $N$ and $m$.  To demonstrate how elongation can be controlled with our method, we choose all of these to have uniform elongation, {\em i.e.} $\rho = \rho_0 \sim 4$-$5$ among these examples.  For low values of $\rho_0$, the limit is approached where the criterion of Eqn.~\ref{eq:rho-crit} is marginally satisfied: Interestingly, the function $\bar{e}$ develops cusp-like behavior at the value of $\varphi$ where $\sigma^2$ approaches $\rho_0^2/4-1$.  The elongation remains smooth (by construction), underscoring the fact that such configurations remain geometrically simple in a sense, despite being potentially difficult to realize with a conventional approach that supplies Fourier coefficients of $\bar{e}$ as input.

\begin{table}
    \centering
\begin{NiceTabular}{c|cccc}
&  $m = 0$ &  $m = 1/2$ &  $m = 1$ &  $m = 3/2$\\
\midrule
     $N=1$ & \includegraphics[width=0.16\textwidth,valign=m]{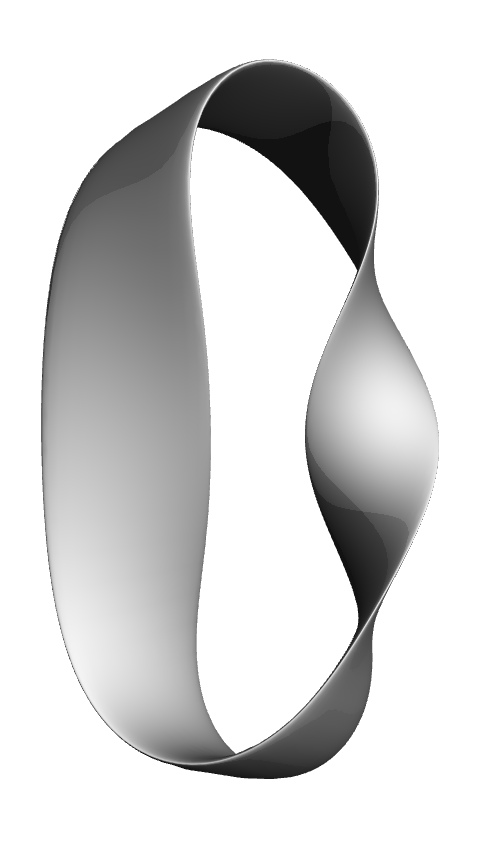}& \includegraphics[width=0.13\textwidth,valign=m]{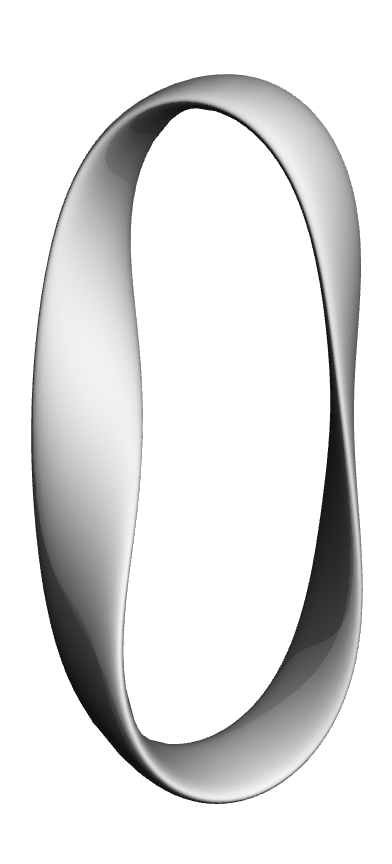} & \includegraphics[width=0.12\textwidth,valign=m]{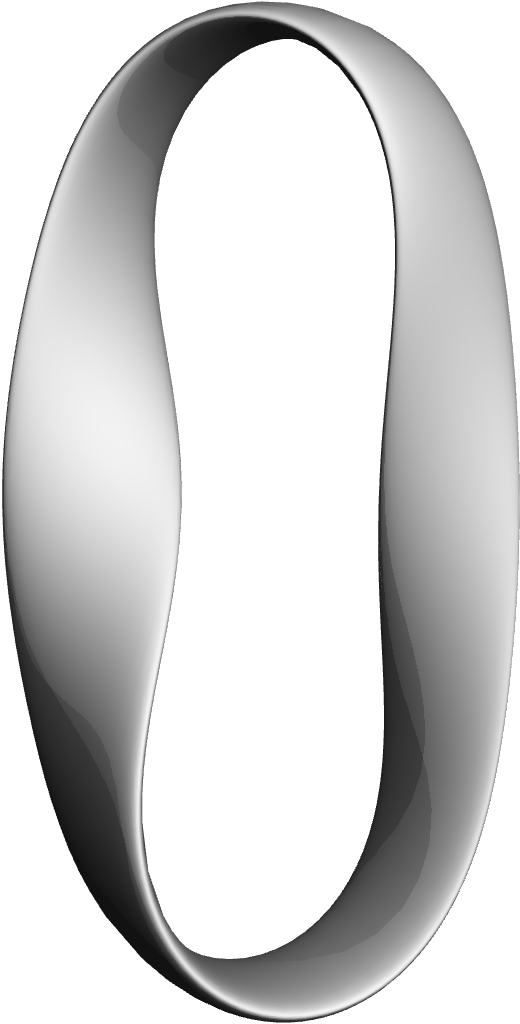} & \includegraphics[width=0.15\textwidth,valign=m]{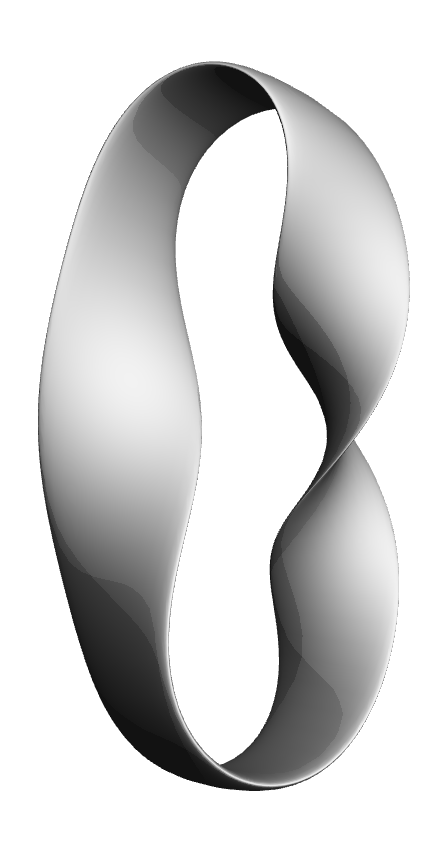}\\
     $N=2$ & \includegraphics[width=0.2\textwidth,valign=m]{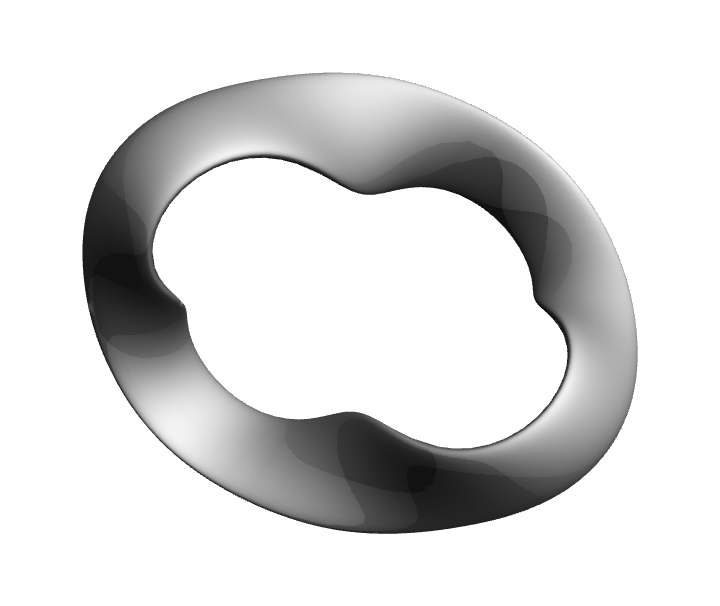}& \includegraphics[width=0.18\textwidth,valign=m]{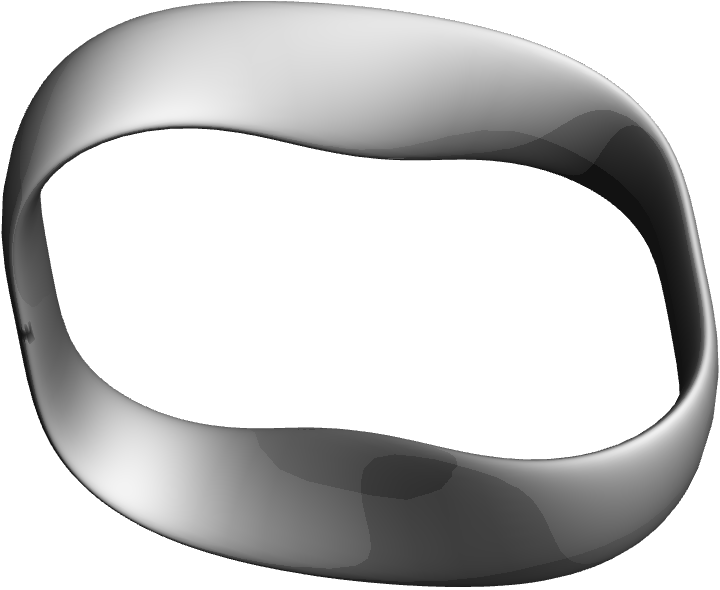} & \includegraphics[width=0.2\textwidth,valign=m]{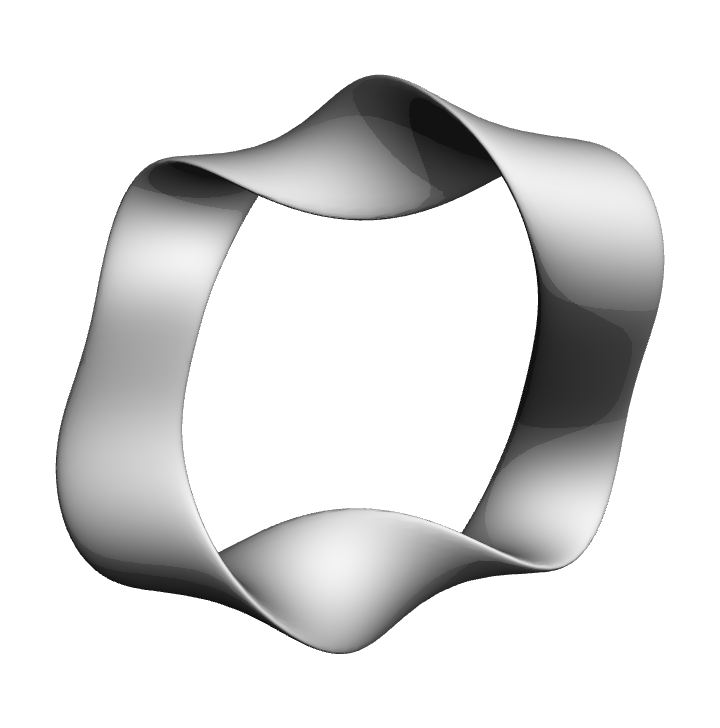} & \includegraphics[width=0.2\textwidth,valign=m]{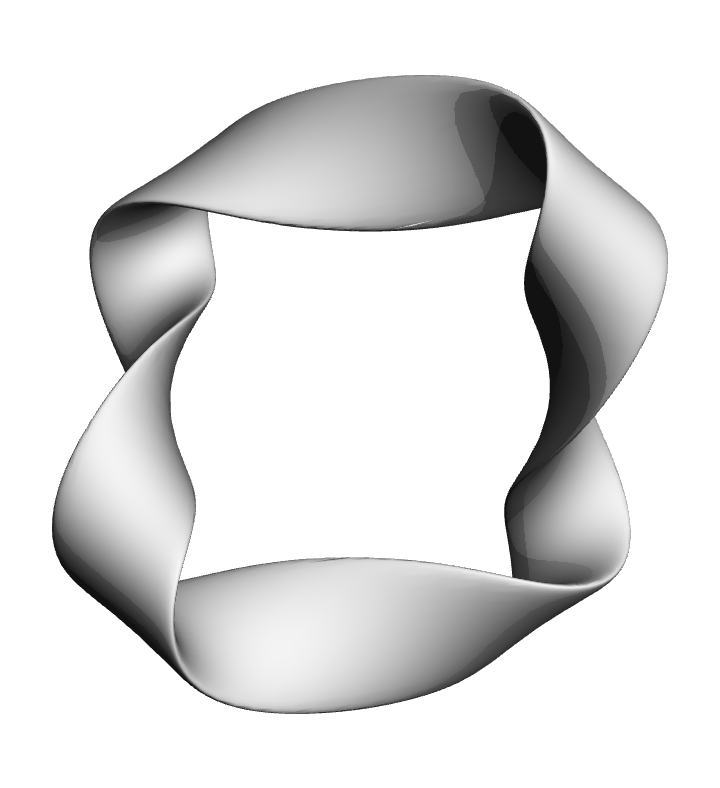}\\
     $N=3$ & \includegraphics[width=0.2\textwidth,valign=m]{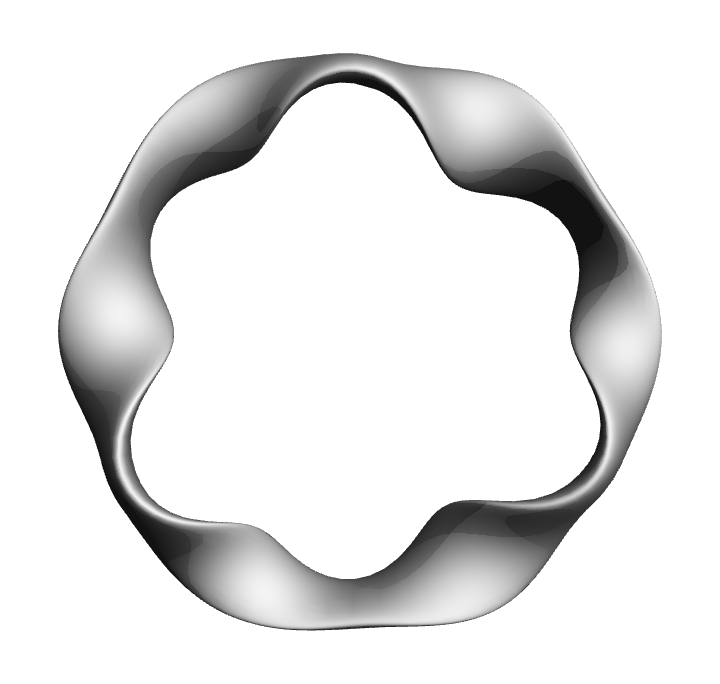}& \includegraphics[width=0.2\textwidth,valign=m]{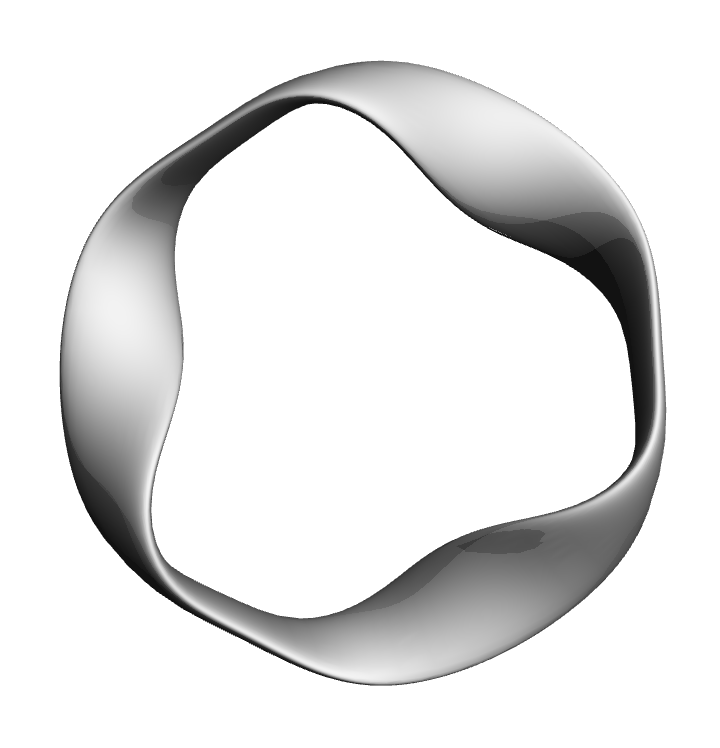} & \includegraphics[width=0.2\textwidth,valign=m]{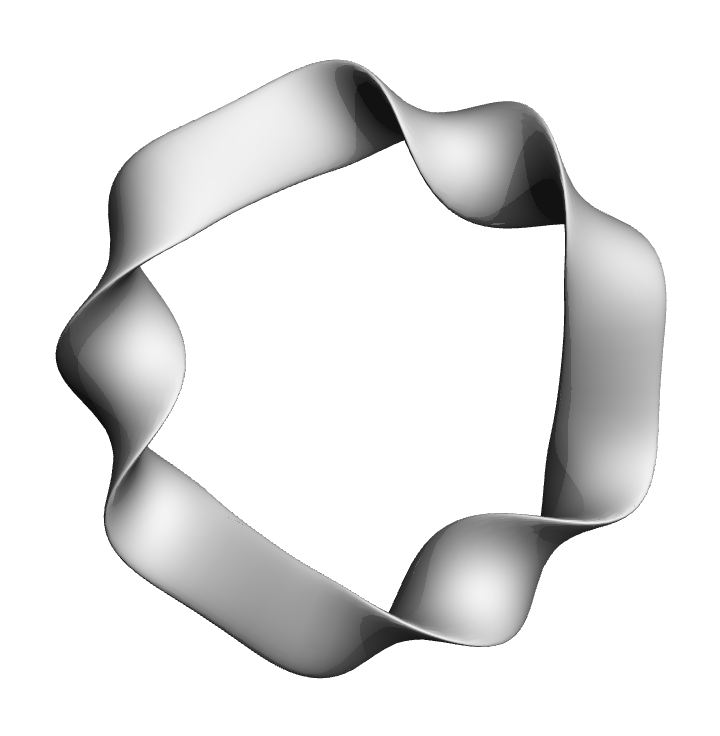} & \includegraphics[width=0.2\textwidth,valign=m]{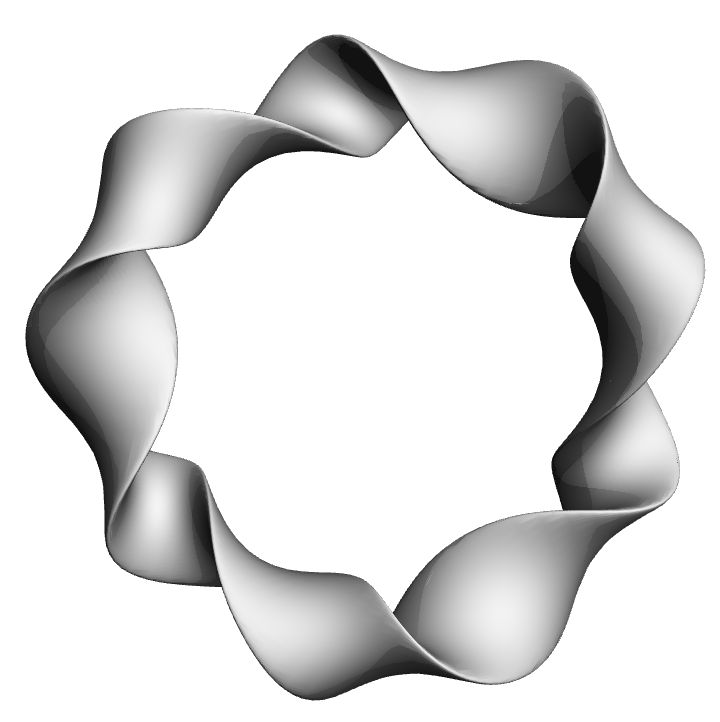}\\
     $N=4$ & \includegraphics[width=0.2\textwidth,valign=m]{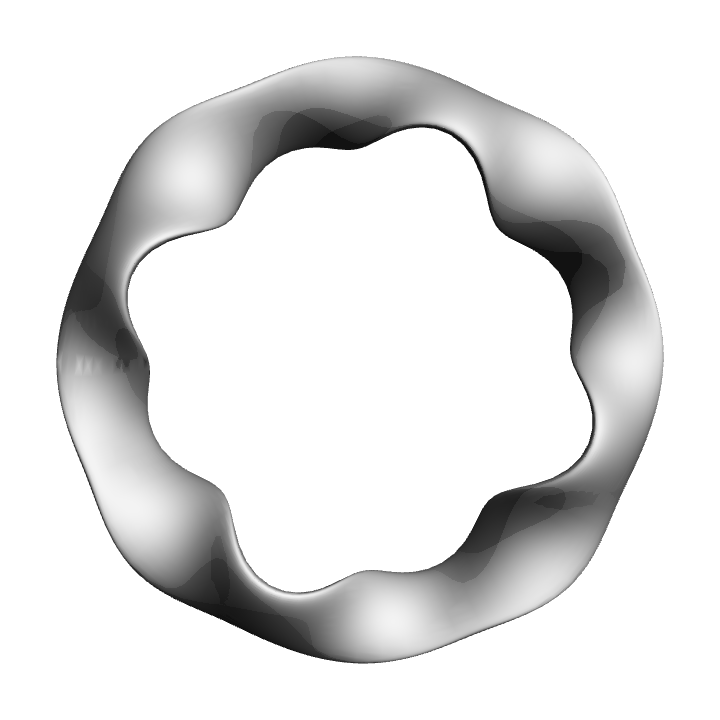}& \includegraphics[width=0.17\textwidth,valign=m]{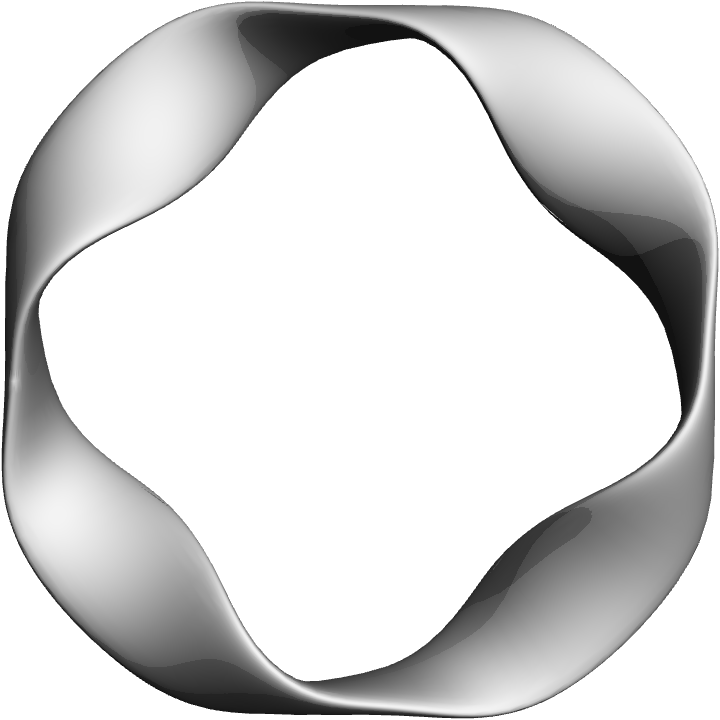} & \includegraphics[width=0.2\textwidth,valign=m]{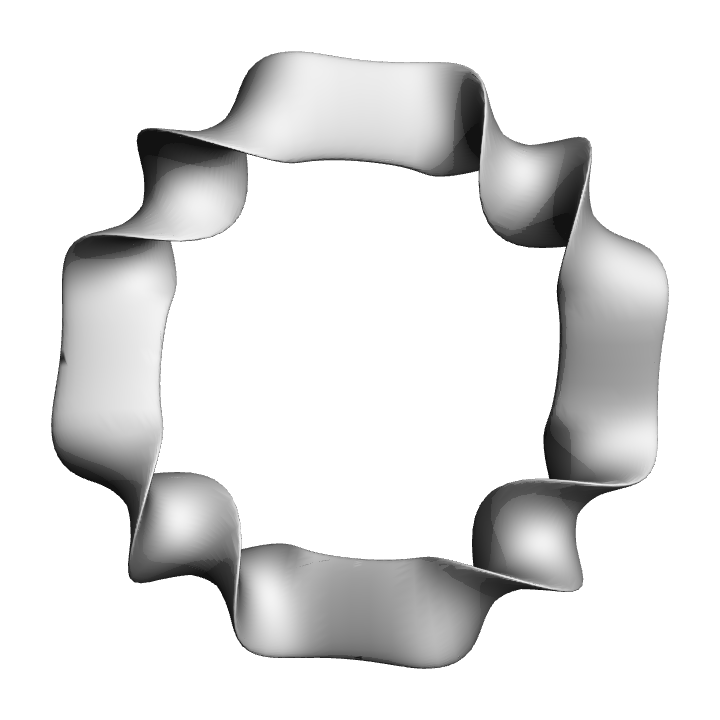} & \includegraphics[width=0.2\textwidth,valign=m]{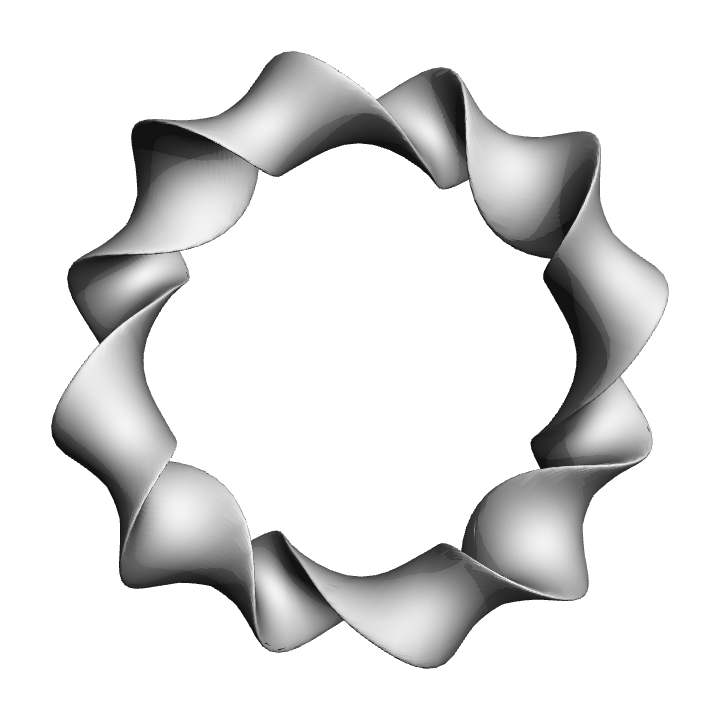}\\
\end{NiceTabular}
    \caption{A view of configurations with varying field period number $N$ and axis helicity $m$.  Noteworthy cases that have been studied previously include $(N, m) = (1, 1)$ \citep{plunk_landreman_helander_2019}, $(N, m) = (2, 1/2)$ \citep{Plunk_figure-8_2025}, and the ever popular choice $(N, m) = (4, 1/2)$ used in present day integrated QI optimization, {\em e.g.} \cite{Goodman_PRXEnergy_2024}.}
    \label{tab:Nm-Grid}
\end{table}

\subsection{Exotic QI}

To further demonstrate the flexibility of the solution method, we show examples of a few QI configuration of the ``knotatron'' type \citep{hudson-knotatron-2014}, where the plasma volume is knotted.  Knotted axis shapes can be found with the method already described.  This may appear disallowed by the assumption \ref{curve-conventions-1}, which would seem to imply conventional axis shapes, where a single period of the plasma equilibrium occupies a sector spanning an interval of $2\pi/N$ in $\phi$.  In actuality assumption \ref{curve-conventions-1} only requires the beginning and end of a field period to span this interval, and makes no restrictions on what happens in between.  This allows axis shapes, for instance, where the curve spans a total angular distance of $-2\pi(N-1)/N$, such that it travels in the negative direction in $\phi$ ($d\phi/d\ell < 0$) and can therefore overlap and link with other field periods of the curve.  The frame at the start and end of a field period can satisfy the closure criteria by being anti-aligned in this case. 

\begin{figure}
    \centering
    \includegraphics[width=0.3\linewidth]{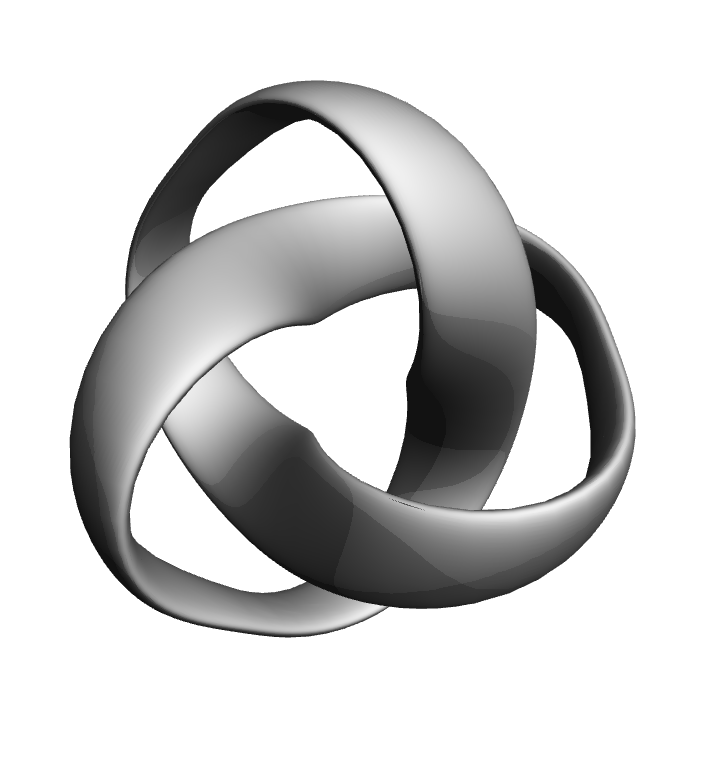}\includegraphics[width=0.3\linewidth]{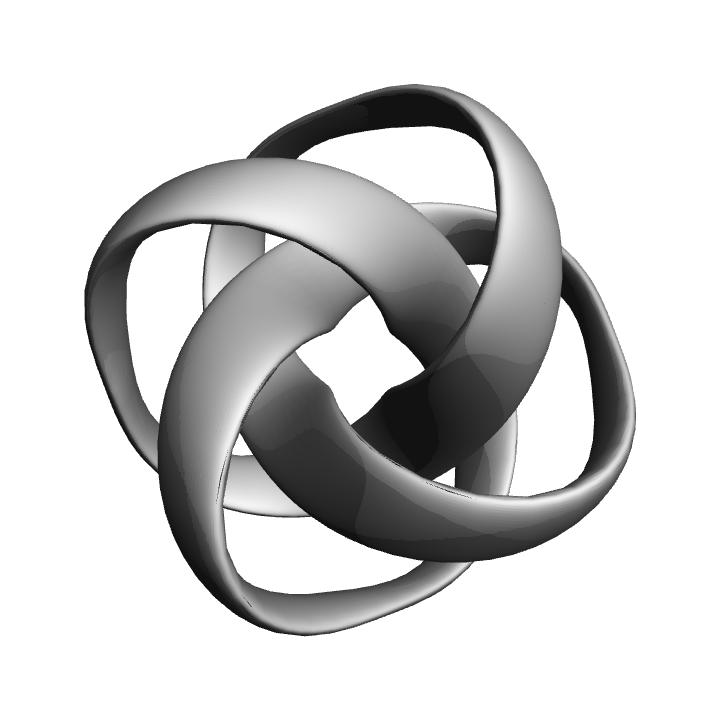}\includegraphics[width=0.3\linewidth]{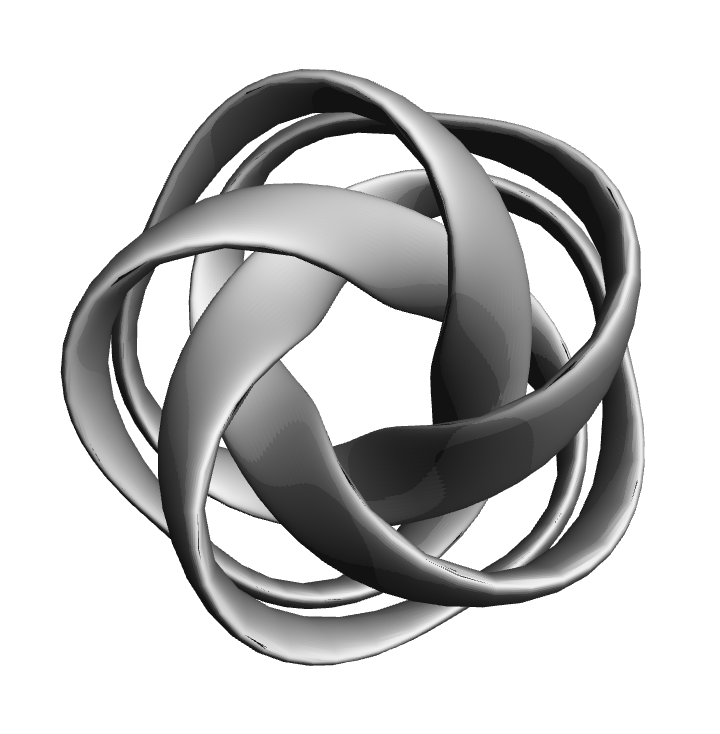}
    \caption{Trefoil, quatrefoil and cinquefoil knotted QI stellarators.}
    \label{fig:knots}
\end{figure}

A concrete example of all this is the $N=3$ trefoil knot, obtained when the axis curve begins at $\phi = 0$, proceeds clockwise and ends at $\phi = 2\pi/3$.  Two more configurations following a similar pattern are found with $N = 4$ and $N = 5$, as also shown in Figure \ref{fig:knots}.   These have increasingly large aspect ratio, taking their effective major radius to be set by the length of the magnetic axis curve, {\em i.e.} $R_\mathrm{eff} = L/(2\pi) = 1$.  Not much is known about such knotted QI configurations, but non-QI knotted stellarators have been shown in the past \cite{garren-boozer-1, hudson-knotatron-2014}, and it is possible now to investigate such configurations beyond near-axis theory with the MHD equilibrium code GVEC \citep{Hindenlang_GVEC-Frenet_2025}.

Note that these configurations have been constructed assuming a somewhat simpler form of curvature that only requires first-order zeros,

\begin{align}\label{eq:kappa-1}
    \kappa =  \kappa_1 \sin(N \ell),
\end{align}
which, for the assumed helicity ($m=1/2$) means that the axis curves are not analytic at field maxima.

It is possible to extend the construction much further than done here, to explore the space of QI stellarators, for instance multiple magnetic trapping wells per field period, or even sub-wells are possible, {\em i.e.} features that arise due to magnetic field maxima distinct from the global maximum \citep{parra-et-al}.  It is also possible to construct QI configurations that break stellarator symmetry, but this will be left for a future paper.

\section{A survey of half-helicity configurations}

Among the large variety of types of QI, there is one which has been much more extensively studied, namely the half-helicity class $m = 1/2$.  This is the class to which the Wendelstein-7X stellarator belongs, as well as most of the modern QI designs found by integrated optimization, {\em e.g.} the SQuIDs of \cite{Goodman_PRXEnergy_2024}.  In this section we make a survey of half-helicity configurations belonging to the class with curvature zeros $(2,3)$.  The on-axis magnetic field is that described by Eqn.~\ref{eq:B0}, with the mirror-ratio parameter set to $\Delta = 0.25$.  This `flat' behavior of $B_0$ near its minimum is fairly ubiquitous with optimized QI designs, and is understood to be conducive to the achievement of a magnetic well with modest shaping at higher order \cite{Rodriguez_Helander_Goodman_2024, Plunk_figure-8_2025}.

\subsection{Space of axis shapes} \label{sec:half-helicity-axes}

Starting with the magnetic axis curve, we will study field periodicity numbers ranging from $N = 2$ to $N = 8$, as we find no interesting qualitative changes arising at larger $N$.  We parameterize this family of curves by the value of the first harmonic of the torsion $\tau_1$; see Eqn.~\ref{eq:tau-N2}.  We generate a large number of axes ($\sim 500$ for each $N$) by varying $\tau_1$ between two extremes where bifurcations occur and the solution branch is lost \citep{rodriguez2023constructing}.  By going to these extremes, we can obtain a comprehensive view of the range of possible shapes, at least for the branches considered.  In principle there may be other branches that have been missed, but we find no evidence of this for $m = 1/2$ and $N \geq 2$.

\begin{figure}
    \centering
    \includegraphics[width=0.44\linewidth]{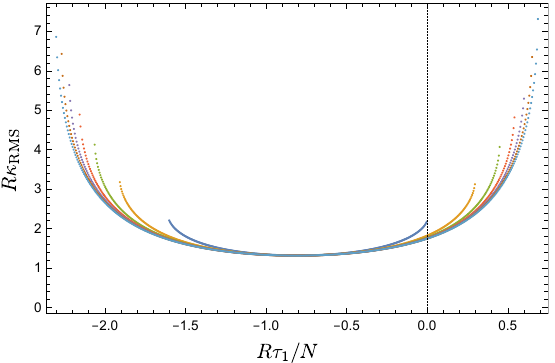}\hfill\includegraphics[width=0.53\linewidth]{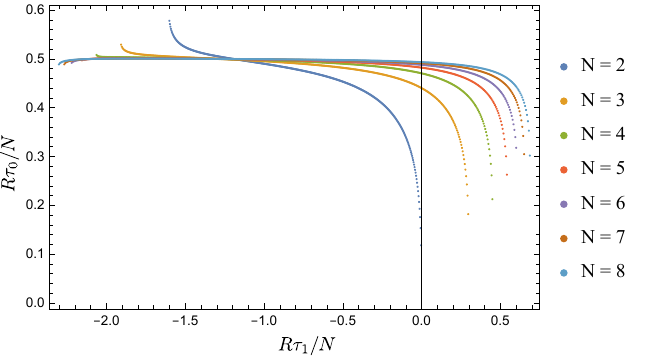}
    \caption{Curvature $\kappa_{\mathrm{RMS}} = \sqrt{3}\kappa_1/8$ and mean torsion $\tau_0$ for family of axis curves parameterized by $\tau_1$.  The effective major radius $R = L/(2\pi)$ is used to normalize both curvature and torsion, where $L$ is the total axis length.}
    \label{fig:axis-parameters}
\end{figure}

\begin{figure}
    \centering
    \includegraphics[width=0.5\linewidth]{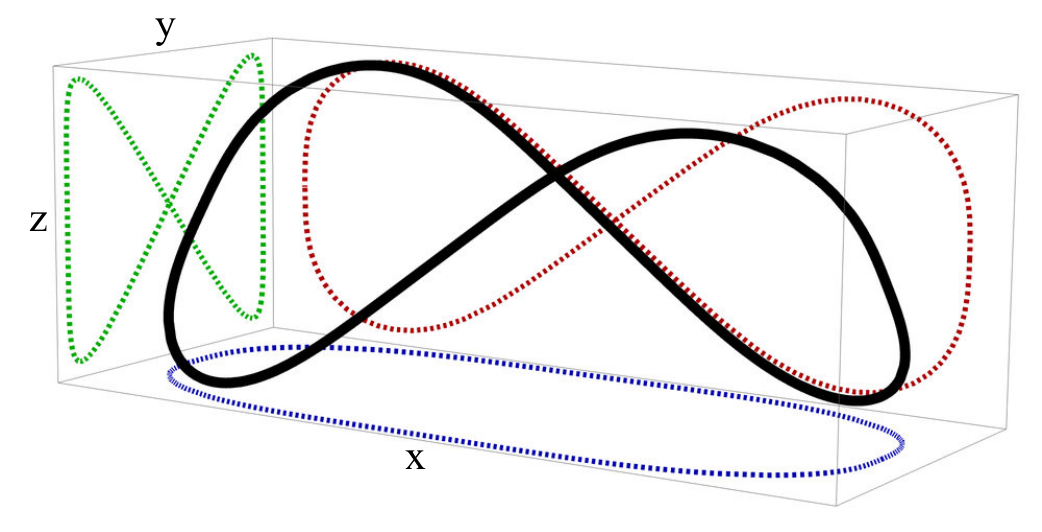}\\
    \caption{Projections of axis shape for $N = 2$, defining ``Bow-tie'' ($y$-$z$), ``Racetrack'' ($x$-$y$) and ``Figure-8'' ($x$-$z$) views.}
    \label{fig:axis-examples}
\end{figure}

\begin{figure}
    \centering
    \includegraphics[width=0.3\linewidth]{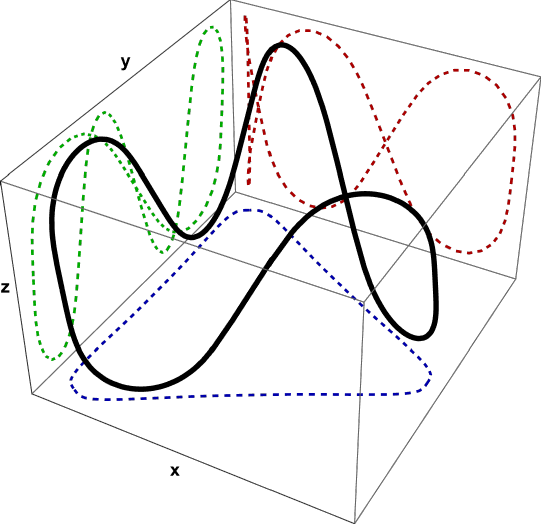}\hspace{0.3cm}\includegraphics[width=0.3\linewidth]{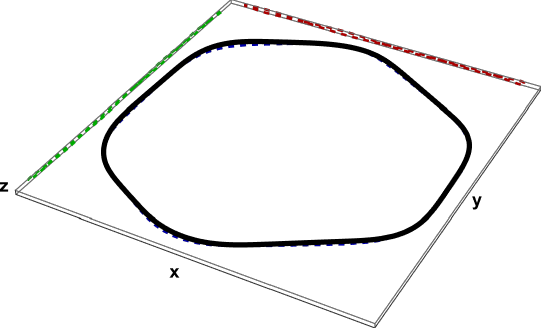}\hspace{0.3cm}\includegraphics[width=0.3\linewidth]{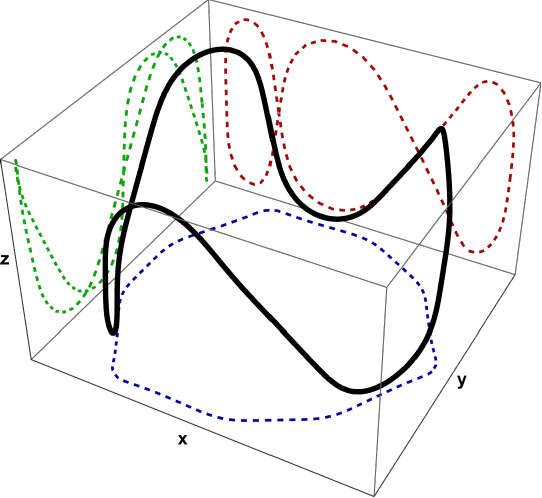}
    \caption{Three extremes of axis shape for the $N = 3$ case, low mean torsion (left; $\tau_1 = 0.89$), low excursion (middle; $\tau_1 = -2.5$), and high mean torsion (right; $\tau_1 = 5.72$).}
    \label{fig:axis-examples-N3}
\end{figure}

\begin{figure}
    \centering
    \includegraphics[width=0.45\linewidth]{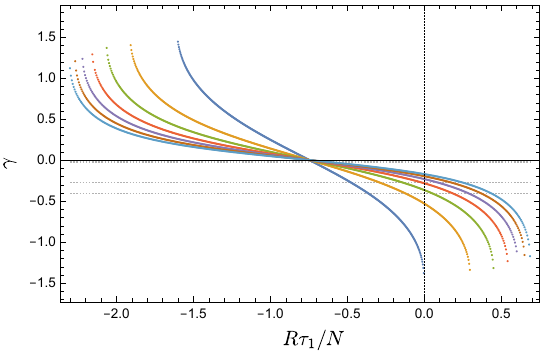}\hfill\includegraphics[width=0.53\linewidth]{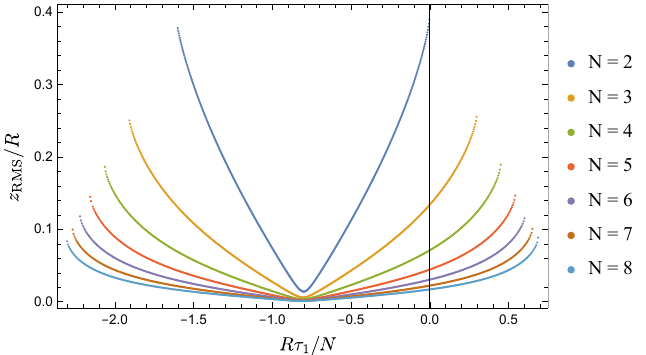}
    \caption{Measures of axis non-planarity, axis inlcination angle $\gamma$ (left) and root-mean-square deviation from the $x$-$y$ plane $z_\mathrm{RMS}$ (right).  The dashed gray horizontal lines (left plot) indicate values of $\gamma$ ($-0.0229$, $-0.270$, $-0.412$) for reference designs W7-AS, W7-X and SQuID-X \citep{Goodman_PRXEnergy_2024}, supporting the idea that larger $|\gamma|$ is compatible with integrated optimization of QI stellarators.}
    \label{fig:axis-non-planarity}
\end{figure}

Figure \ref{fig:axis-parameters} gives an overview of the parameters found for the entire set of axis curves.  The value of $\tau_1$ is normalized by $N$, as torsion is observed to increase with $N$.  Axes with $N = 2$ ({\em e.g.} Fig.~\ref{fig:axis-examples}) show special behavior as $\tau_1$ approaches $0$, namely that overall torsion tends to zero.  This is the planar figure-8 limit recently studied in \cite{Plunk_figure-8_2025}.  Figure \ref{fig:axis-examples} and Table \ref{tab:figure8-sequence} give an overview of the geometry of this class of curves, including the high-mean-torsion limit neglected in Ref.~\citep{Plunk_figure-8_2025}.  The space of axes for $N \geq 3$ have qualitatively similar behavior for the different field period numbers.  In one extreme (largest $\tau_1$) there is low mean torsion but relatively high axis curvature.  In this extreme field period numbers $N \geq 3$ cannot collapse into a plane as with the case of the $N = 2$ figure-8, but instead have a ``tall'' crown-like appearance, {\em i.e.} have a large extent in the $z$ direction; see also Figure \ref{fig:axis-non-planarity}.  These are the high ``writhe'' \citep{fuller-writhe} analogues of the figure-8, which manage to generate axis helicity with relatively little mean torsion  \citep{Plunk_figure-8_2025}.  At the opposite extreme (negative $\tau_1$), the curves also extend strongly out of the $x$-$y$ plane, but at the cost of large mean torsion, {\em i.e.} the torsion seems to be at odds with the helicity.  Example axis shapes demonstrating these limits are shown in Figure \ref{fig:axis-examples-N3} for $N = 3$.

\begin{table}
\centering
\begin{NiceTabular}{c|c|c|c}
  &  low mean torsion & low inclination & high mean torsion  \\
\midrule
Bow-tie view  &  \includegraphics[width=0.15\linewidth, valign=m]{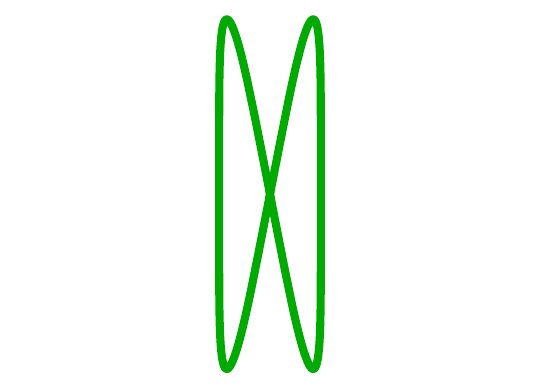} & \includegraphics[width=0.15\linewidth, valign=m]{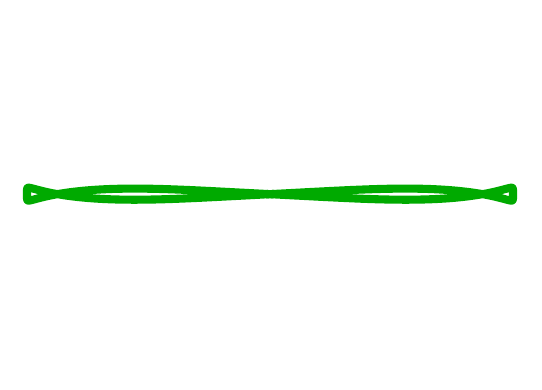} & \includegraphics[width=0.15\linewidth, valign=m]{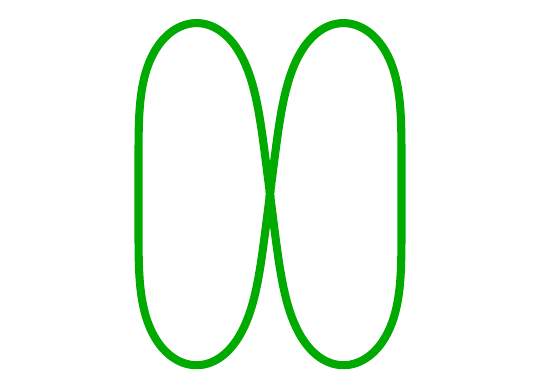}\\
Racetrack view  &  \includegraphics[width=0.15\linewidth, valign=m]{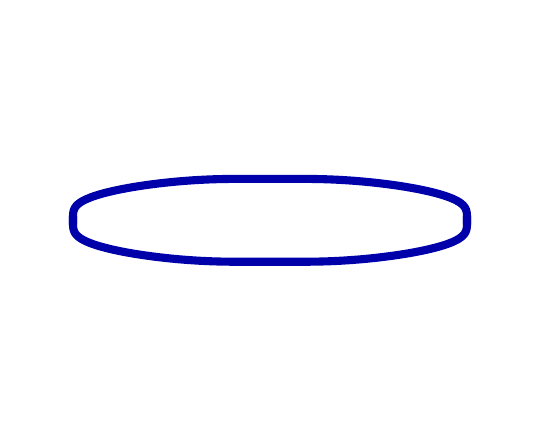} & \includegraphics[width=0.15\linewidth, valign=m]{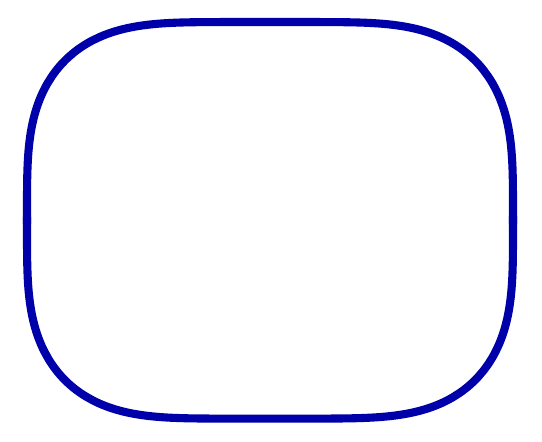} & \includegraphics[width=0.15\linewidth, valign=m]{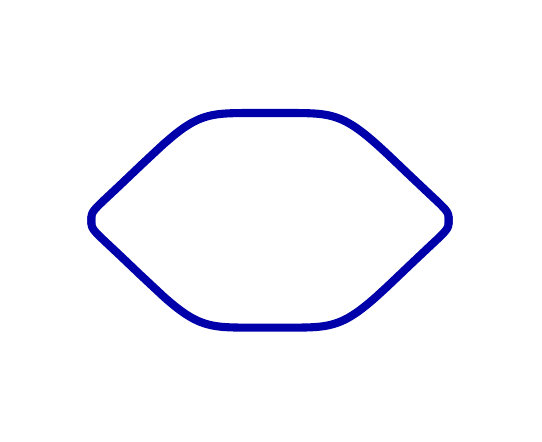}\\
Figure-8 view  &  \includegraphics[width=0.15\linewidth, valign=m]{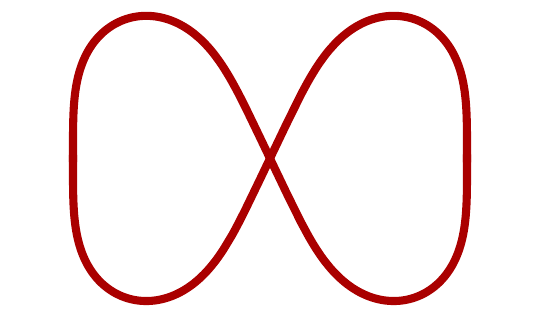} & \includegraphics[width=0.15\linewidth, valign=m]{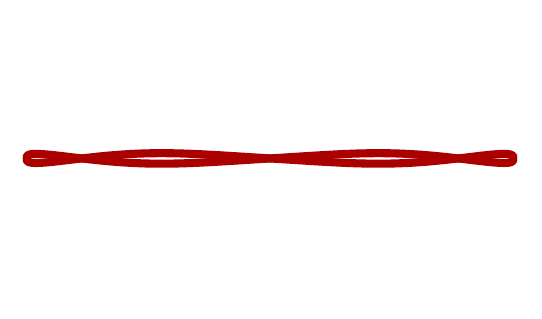} & \includegraphics[width=0.15\linewidth, valign=m]{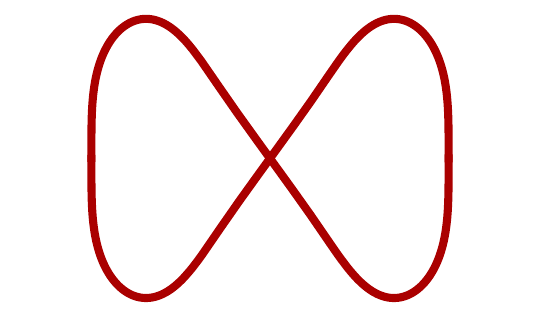}\\
\end{NiceTabular}
\caption{Three views of three axis shapes, demonstrating range of behavior within the curve family: high writhe, low mean torsion (left), low excursion in $z$ (center), and low writhe, high mean torsion (right).  $N=2$ is chosen for simplicity, and projections are explained by Figure \ref{fig:axis-examples}}
\label{tab:figure8-sequence}
\end{table}

For a more quantitative picture of axis non-planarity, we plot two further quantities in Figure \ref{fig:axis-non-planarity}.  The two plots show the axis inclination angle $\gamma$, and the root-mean-square excursion from the $x$-$y$ plane $z_\mathrm{RMS}$, defined according to

\begin{eqnarray}
    \gamma = \arctan(\t\cdot\hat{\bf \phi}|_{\phi = 0}, \t\cdot\hat{\bf z}|_{\phi = 0}),\\
    z_\mathrm{RMS} = \sqrt{\frac{\pi}{N}\int_0^{\pi/N} |{\bf x}_0\cdot\hat{\bf z}|^2 d\ell},
\end{eqnarray}
with $\arctan(x,y)$ the two-argument arctangent.  The inclination angle $\gamma$ was originally used by Spitzer to characterize non-planarity \citep{spitzer-1958, Plunk_figure-8_2025}: a $|\gamma|=\pi/2$ would correspond to a completely vertical section of the axis.  We note that only the magnitude of $\gamma$ was reported in \cite{Plunk_figure-8_2025} for $N = 2$, while here we include the sign, and vary over the full range of the input parameter $\tau_1$ to show opposite extremes of non-planarity, including high and low mean torsion.  We observe that although the excursion does reduce in $N$, the inclination angle can still approach extreme values $-\pi/2$ and $\pi/2$.  Note that such extreme axis shapes preclude the use of cylindrical coordinates, underscoring the potential value of more flexible approaches to solving the equilibrium problem \cite{Hindenlang_GVEC-Frenet_2025}.

For all $N$, there is also a unique axis shape of minimal curvature that occurs at an intermediate value of $\tau_1$.  These curves lie close to the $x$-$y$ plane, {\em i.e.} they are also cases of minimal excursion, $z_\mathrm{RMS}$.  Remarkably, all the axis parameters (including $\tau_0/N$) line up almost perfectly for the case of minimal curvature.  Therefore a scaling symmetry is approximately satisfied (assuming equal total axis lengths) for the curvature and torsion, where the functions $\kappa(\hat{\ell}/N)$ and $\tau(\hat{\ell}/N)/N$ are nearly equal for all $N \geq 2$ and all $\hat{\ell}$.

It is not clear whether this apparent symmetry is a special property of the family of axis curves considered, or if it is more universally true, but we have confirmed it for other axis families (choices of the orders of zeroes of curvature). As can be seen in Figure \ref{fig:axis-parameters} there is a rather broad region in parameter space (especially for $N \geq 3$) over which the scaling symmetry is approximately satisfied.  That is, the axis parameters $\kappa_1$ and $\tau_0/N$ are close in magnitude for a range of values $\tau_1$ about the case of minimum curvature.  As we will see in the following section, the consequences of this approximate scaling symmetry of the axis shape extend to the full near-axis solution of the QI field.

\subsection{First order solutions}

Configurations are generated by varying axis and shaping parameters, holding axis helicity $m$ and field periodicity $N$ constant.  This allows a specific class of solutions to be studied in detail.  By sweeping over geometric parameters, a database of configurations can be generated.  A difficulty arises in that it is not known {\it a priori} which values of the input parameters will yield valid solutions, for example what limits are placed on $\tau_1$ as shown in Figure \ref{fig:axis-parameters}, and what values of $\rho_0$ and $\rho_1$ are consistent with valid (real and periodic) solutions for $\sigma$.  Thus the parameters must be scanned methodically to find the extremities of parameter space.  A set of configurations was generated in this manner, for $N=2$-$6$, varying elongation parameters within $\rho_0 \in [2.4, 4.5]$ and $\rho_1 \in [-1.5, 1.5]$, and varying the axis parameter $\tau_1$ between the case of minimal curvature (and inclination), and the extreme case of minimal mean torsion (and high inclination); see Figures \ref{fig:axis-parameters} and \ref{fig:axis-non-planarity}. Such databases of configurations have been used to study QI configurations in other recent works \citep{Rodriguez_Plunk_Residual_2024, rodriguez2025measures}.

For all the configurations obtained, the outer surface shape was used as input for the \texttt{VMEC} code \citep{Hirshman}.  Many of the \texttt{VMEC} runs fail to initialize or converge, especially for configurations with large inclination angles $|\gamma|$ due to the difficulties associated to the use of cylindrical coordinates, and such results were discarded.

The sample presented here is only a small one. A significantly broader parameter scan will be investigated in a future publication, using more sophisticated tools based on higher order near-axis theory \citep{Rodriguez_Plunk_2023_Higher_QI,Rodríguez_Plunk_Jorge_2025,rodriguez2025measures}.

\subsection{$\delta\! B$ and scaling with $N$}

The approximate scaling symmetry of the axis shapes, detailed in Section \ref{sec:half-helicity-axes}, implies a related symmetry for the full first-order solution.  In particular the $\sigma$-equation \ref{eq:sigma} can be transformed using

\begin{eqnarray}
    \hat{\varphi} &= N \varphi,\\
    \hat{\tau} &= \tau / N,\\
    \hat{\iotaslash}_0 &= \iotaslash_0 / N,\\
    \hat{\sigma}(\hat{\varphi}) &= \sigma(\varphi),
\end{eqnarray}
leaving the equation written in hatted quantities unchanged.  Therefore the function $\sigma(\hat{\varphi})$ is independent of $N$ to the extent that symmetry is satisfied by the axis shape ({\em e.g.} $\kappa(\hat{\ell}/N)$ and $\tau(\hat{\ell}/N)/N$ do not vary significantly with $N$ or $\hat{\ell}$).  Because $\kappa$ is preserved, so is the field strength, and therefore the full magnetic field obeys a symmetry as well.

\begin{figure}
    \centering
    \includegraphics[width=0.49\linewidth]{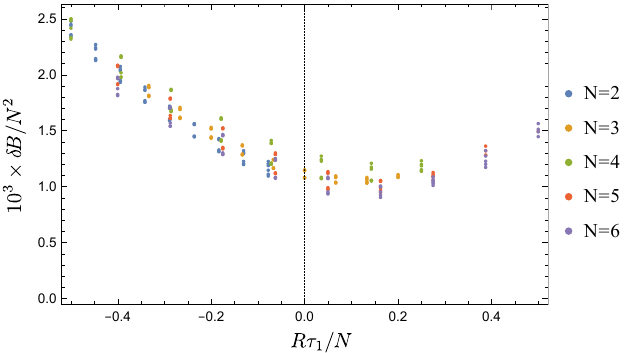}\hfill\includegraphics[width=0.47\linewidth]{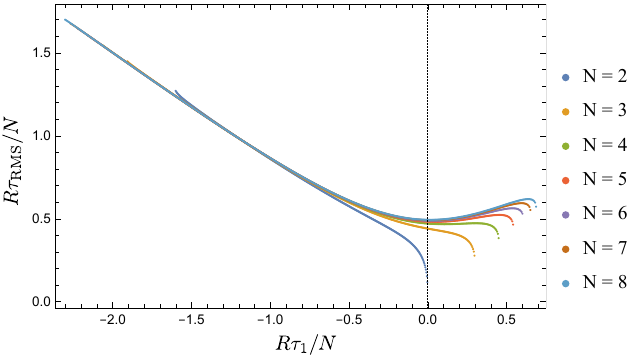}
    \caption{Scaling symmetry of equivalently shaped QI configurations (left); see Eqn.~\ref{eq:Qi-scaling-law}.  On the right is another view of axis parameters, showing some correlation between quality $\delta\!B$ and the size of axis torsion $\tau_\mathrm{RMS}$.}
    \label{fig:deltaB-scaling}
\end{figure}

It should not be too surprising that the symmetry should extend to higher order in the NAE, since the first order solution enters as input to the second order theory, and so forth.  To see evidence of this, we can evaluate the root-mean-square deviation in the magnetic field strength of constructed configuration, evaluated with the \texttt{VMEC} equilibrium code, from the theoretical first order form \citep[Eq.~(3.1)]{rodriguez2025measures}:

\begin{equation}
    \delta\!B=\sqrt{\int(B_\texttt{vmec}-B_\mathrm{nae}^{1\mathrm{st}})^2\mathrm{d}\varphi\,\mathrm{d}\theta}\Bigg/\sqrt{\int B_\texttt{vmec}^2\mathrm{d}\varphi\,\mathrm{d}\theta}. \label{eqn:dB}
\end{equation}
This quantity, which measures the importance of higher order corrections, is plotted in Figure \ref{fig:deltaB-scaling} for a number of configurations from the database, with approximately the same uniform elongation profile $|\rho - 4| < 0.35$.  Scanning with respect to the axis parameter $\tau_1$, we see that $\delta\!B/N^2$ is approximately independent of $N$.  It is also true asymptotically that $\delta\!B \sim (a/R)^2$ where $a/R = \sqrt{2}\epsilon$ is the inverse aspect ratio written with the conventions of this paper.  The confirmation of aspect ratio scaling, which is not shown here, merely confirms that the NAE was performed correctly.  (Note that all solutions shown in Figure \ref{fig:deltaB-scaling} have aspect ratio $R/a = 14.14$.)  Thus we arrive at the following scaling law

\begin{equation}
     \delta\!B \approx C_g \left(\frac{aN}{R}\right)^2,\label{eq:Qi-scaling-law}
\end{equation}
where $C_g$ is a dimensionless geometric constant that depends on field shaping parameters ($\rho_0$, $\tau_1$, {\em etc.}), but is largely independent of aspect ratio $R/a$ and field period number $N$ for $N \geq 2$.  Although an investigation of detailed geometric dependence of $C_g$ is beyond the scope of this paper, it is plausible that the torsion is a key control parameter, as indicated by previous work \citep{camacho-mata_plunk_2022}; note the qualitative similarities between the plots of $\delta\!B$ and $\tau_\mathrm{RMS}$ as shown in Figure \ref{fig:deltaB-scaling}.

The scaling of Eqn.~\ref{eq:Qi-scaling-law} suggests an underlying reason why low field period configurations can be designed more compactly, since the truncation error (and higher order influence on the magnetic field that potentially spoil QI quality) can be controlled at low aspect ratio if $N$ is also low.  We stress however that this comparison, made with \texttt{VMEC}, was limited to configurations with relatively weak non-planarity, and therefore excludes those axes (values of $\tau_1$) for which the scaling symmetry is more strongly violated; see Figures \ref{fig:axis-parameters} and \ref{fig:axis-non-planarity}.  Thus we expect the scaling \ref{eq:Qi-scaling-law} to apply only to QI stellarators that are sufficiently weakly non-planar.

\subsection{The curious case of $N = 1$}

For $N = 1$ we can similarly vary shaping parameters to explore the geometry of the magnetic axis and the space of QI configurations.  Although the axis parameter space is different, requiring more parameters to ensure closure, the main qualitative features remain, {\em i.e.} that the space has extremes of high and low mean torsion, between which there lies a case of minimal axis curvature; see Table \ref{tab:N1-sequence}.  In fact, the value of the minimum curvature, $\kappa_\mathrm{RMS} \approx 1.309$ is quite close to the values found for $N \geq 2$; for instance for $N=2$ the value is $\kappa_\mathrm{RMS} \approx 1.316$.

\begin{table}
\centering
\begin{NiceTabular}{c|c|c|c}
  &  low mean torsion & low inclination & high mean torsion  \\
\midrule
$y$-$z$ Projection  &  \includegraphics[width=0.15\linewidth, valign=m]{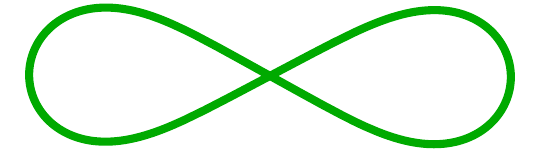} & \includegraphics[width=0.15\linewidth, valign=m]{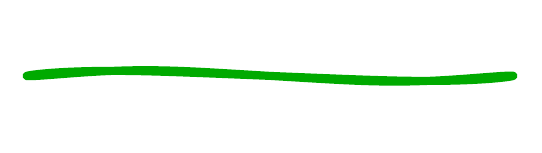} & \includegraphics[width=0.15\linewidth, valign=m]{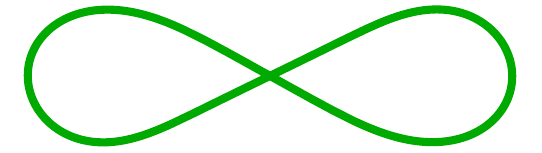}\\
$x$-$y$ Projection  &  \includegraphics[width=0.15\linewidth, valign=m]{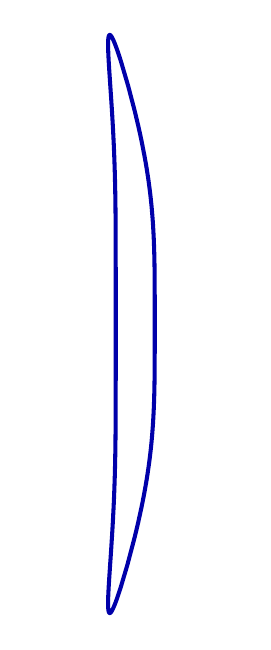} & \includegraphics[width=0.15\linewidth, valign=m]{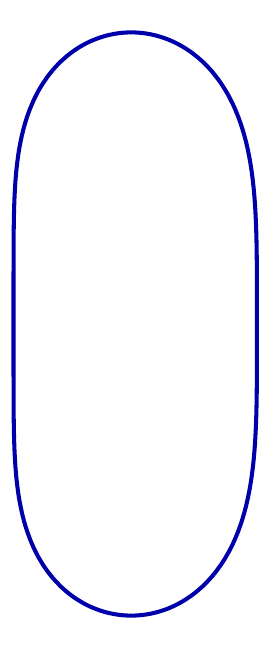} & \includegraphics[width=0.15\linewidth, valign=m]{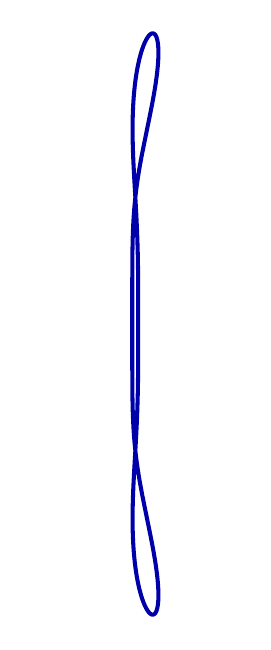}\\
$x$-$z$ Projection  &  \includegraphics[width=0.15\linewidth, valign=m]{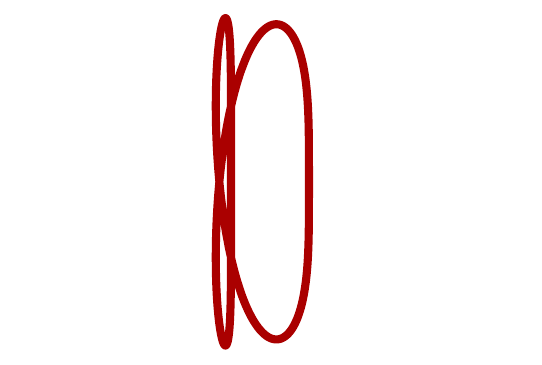} & \includegraphics[width=0.15\linewidth, valign=m]{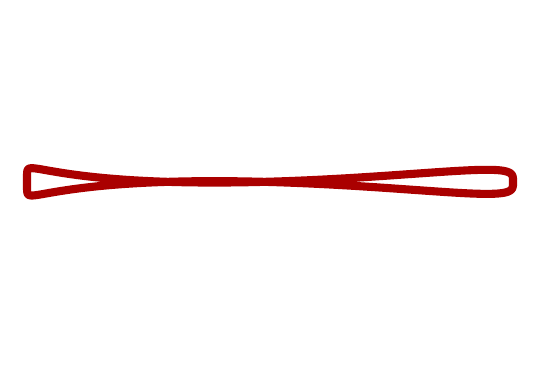} & \includegraphics[width=0.15\linewidth, valign=m]{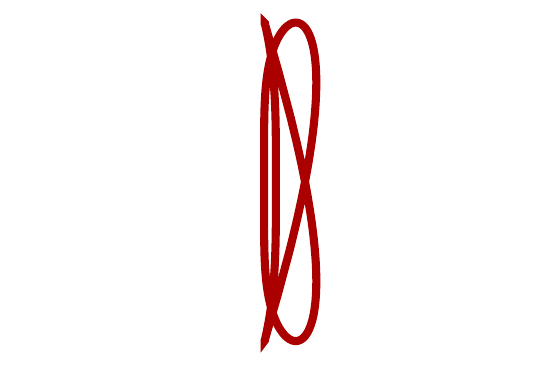}\\
\end{NiceTabular}
\caption{Three views of three axis shapes for $N = 1$: high writhe, low mean torsion (left), low excursion in $z$ (center), and low writhe, high mean torsion (right).  Projections are explained by Figure \ref{fig:axis-examples}}
\label{tab:N1-sequence}
\end{table}

\begin{figure}
    \centering
    \includegraphics[width=0.48\linewidth]{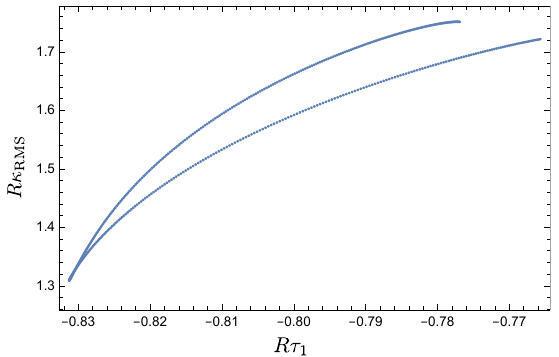}\hfill\includegraphics[width=0.48\linewidth]{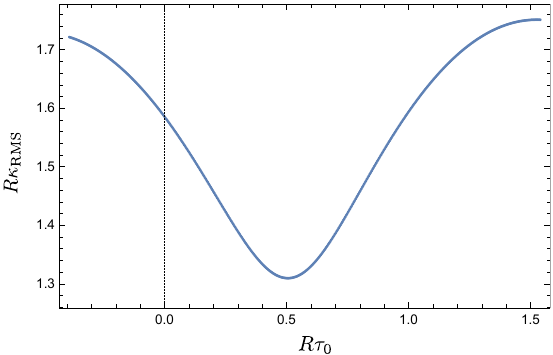}
    \caption{Curvature $\kappa_{\mathrm{RMS}} = \sqrt{3 \kappa_1^2 + 2 \kappa_1 \kappa_2 + 2 \kappa_2^2}/8$ for family of axis curves with $N=1$ parameterized by $\tau_0$.  The effective major radius $R = L/(2\pi)$ is used to normalize both curvature and torsion, where $L$ is the total axis length.}
    \label{fig:axis-parameters-N1}
\end{figure}

Despite these similarities, there are significant differences across the space, like a much larger variation in $\tau_0$ compared to that of $\tau_0/N$ for $N \geq 2$, and much smaller variation of $\tau_1$.  There are also multiple branches when parameterized by $\tau_1$, which is why we use $\tau_0$ instead as the input parameter for these calculations; see Figure \ref{fig:axis-parameters-N1}.  Note also that the case of zero mean torsion (center column of Table \ref{tab:N1-sequence}) is not very inclined, while the case of strongest negative mean torsion (left column of Table \ref{tab:N1-sequence}; $\tau_0 \approx -0.39$) does not tend to a shape fully contained in the $x$-$z$ plane, as with the $N=2$ figure-8 configuration.

\begin{figure}
    \centering
    \includegraphics[width=1.0\linewidth]{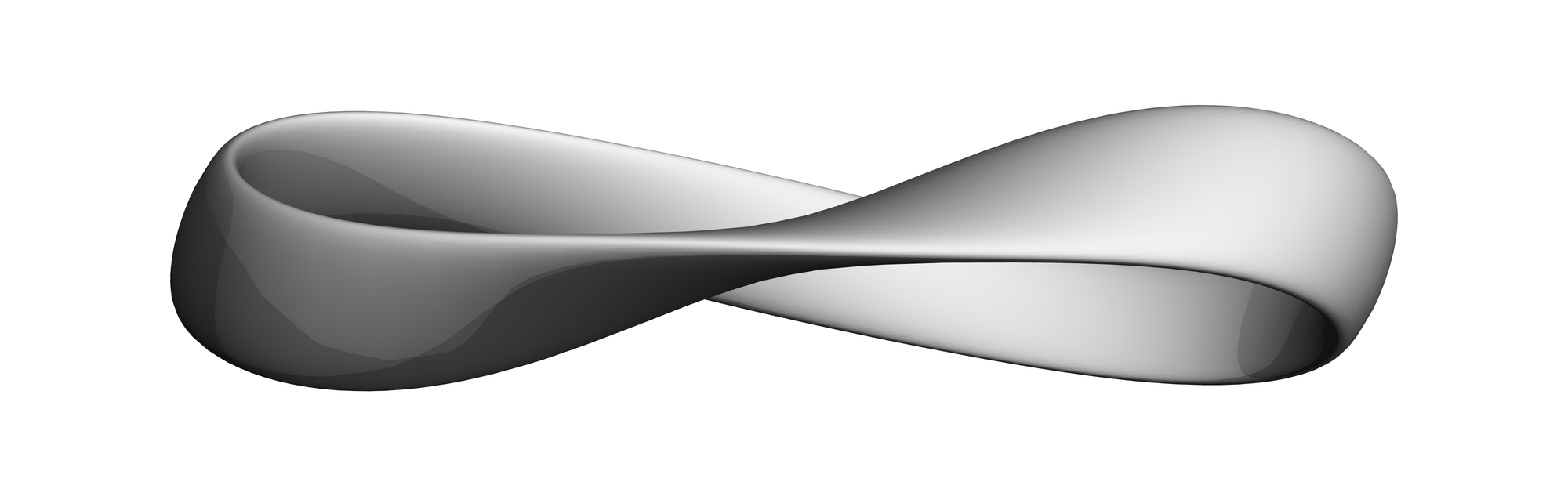}
    \caption{The M\"{o}bius stellarator: a single field period QI stellarator with axis helicity $m = 1/2$.  Its shaping parameters are $\rho_0 =4$, $\rho_1=0$, $\tau_0 = 0.3$ and $\epsilon = .07$ ($R/a \sim 10$).}
    \label{fig:N1-example}
\end{figure}

It is not clear if the $N=1$ configurations offer significant advantages -- note that their $\delta\!B$ scaling is worse than that for $N \geq 2$, {\em i.e.} Eqn.~\ref{eq:Qi-scaling-law}.  Comparing configurations with axes of minimal $\kappa_\mathrm{RMS}$, aspect ratio, and shaping parameters ($\rho_n$), the absolute size of $\delta\!B$ lies between those values obtained for $N=2$ and $N=3$.  It is therefore not expected that the most compact designs possible will have $N=1$.  The larger distances along the field line between neighboring field periods also affects particle orbit widths, negatively affecting things like zonal flows, turbulence and Shafranov shifts \citep{Plunk_Helander_Residual_2024, Rodriguez_Plunk_Residual_2024, rodriguez2025measures}.  However, the ultimate evaluation of such configurations must be done via integrated optimization, which takes a large number of criteria into account simultaneously.  Even if the $N=1$ configurations do not turn out to be of major practical importance, they have a certain fascination and charm; see Figure \ref{fig:N1-example}.

\section{Conclusion}

In this paper we have presented a method to solve for near-axis QI fields to first order, with the inputs reformulated in geometric terms.  Motivated by a strong sensitivity of QI fields to both torsion and curvature, we propose a method to construct the axis curve (zeroth order theory) by solving the Frenet-Serret equations.  This allows for precise control of torsion and curvature, which we demonstrate is more difficult to achieve via the conventional approach of supplying Fourier modes of an axis curve in cylindrical coordinates.

At first order, the near-axis problem is reformulated with elongation as one of the inputs.  This approach, which is related to Mercier's original axis expansion, gives direct control over a feature of central importance in QI optimization that has been found particularly difficult to control with the near-axis approach.

The first-order part of our method is limited to stellarator symmetric quasi-isdodynamic fields because the condition for enforcing QI can be stated simply in terms of a symmetry on the elongation profile.  Despite this limitation, we note that there is a large and varied space of stellarators contained in this class, thanks both to the large dimensionality of QI fields, as compared with QS fields, and the different possible classes of helicity.

Exploration of this space is ongoing and will be the subject of future publications.  Joining the methodology presented in this paper with the techniques and measures recently devised for constructing QI fields to second order has already begun \citep{Rodríguez_Plunk_Jorge_2025, rodriguez2025measures}.  Further development of this kind will enable much broader exploration of parameter space, and the exploration of fundamental tradeoffs involved in the integrated optimization of QI stellarators.

\section*{Acknowledgments}  The authors are grateful for many conversations with Katia Camacho Mata.  This work has been carried out within the framework of the EUROfusion Consortium, funded by the European Union via the Euratom Research and Training Programme (Grant Agreement No 101052200 — EUROfusion).  Views and opinions expressed are however those of the author(s) only and do not necessarily reflect those of the European Union or the European Commission.  Neither the European Union nor the European Commission can be held responsible for them.

\appendix

\section{Describing the axis torsion} \label{sec:app_torsion}

Let us consider a {simple} (non-intersecting) regular (smooth) closed curve, and specialise, for simplicity, on those that may be parametrised continuously by a cylindrical angle. A detailed description of such curves, and in particular ones with flattening points, can be found in Appendix~A of \cite{Rodríguez_Plunk_Jorge_2025}. The embedded curve in $\mathbb{R}^3$ is described by,
\begin{equation}
    \mathbf{x}_{0} = \mathbf{r}_\mathrm{axis}(\phi)=R(\phi)\hat{\pmb R}(\phi)+Z(\phi)\hat{\pmb z}, \label{eqn:R_and_Z}
\end{equation}
where $\{\hat{\pmb R}(\phi),\hat{\pmb \phi}(\phi), \hat{ \pmb z}\}$ represent the orthonormal basis of the cylindrical coordinates $\{R,\phi,z\}$. By considering the functions $R$ and $Z$ to be $C^\infty$ and $2\pi$-periodic in $\phi$, we guarantee the curve to be closed and regular.  In addition, we assume the curve to be stellarator symmetric \citep{dewar1998stellarator} about a point $\phi_0$, if $R(\phi_0+\hat{\phi})=R(\phi_0-\hat{\phi})$ and $Z(\phi_0+\hat{\phi})=-Z(\phi_0-\hat{\phi})$. That is, $R$ is even and $Z$ is odd about \textit{stellarator-symmetric points}. 
\par
To make this direct description of the curve applicable to QI fields we must have a way of imposing certain behavior on curvature and torsion, first and foremost being the requirement of points of zero curvature.

\subsection{Curvature}
To impose local curvature constraints through the functions $R(\phi)$ and $Z(\phi)$, we consider a local description of the neighborhood of a point $p$ on the curve. We follow the approach of \cite{camacho-mata_plunk_2022} (also reminiscent of the so-called local canonical form \citep[Ch.~1.6]{do2016differential}), but extend it to as general terms as possible. 
\par
We define the curvature of the curve as follows. First, we may compute the tangent to the curve $ \hat{\pmb t}=\mathrm{d}\mathbf{r}_\mathrm{axis}/\mathrm{d}\ell$ and its derivative, to construct 
\begin{equation}
    \mathbf{v}=\mathbf{r}_\mathrm{axis}'\times\mathbf{r}_\mathrm{axis}''=\begin{pmatrix} R' \\ R \\ Z' \end{pmatrix}_\mathrm{cyl}\times\begin{pmatrix} R''-R \\ 2R' \\ Z'' \end{pmatrix}_\mathrm{cyl}
    =\begin{pmatrix} RZ''-2R'Z' \\ Z'(R''-R)-R'Z'' \\ 2(R')^2+R(R-R'') \end{pmatrix}_\mathrm{cyl} \; : \;
    \begin{pmatrix} \mathrm{O} \\ \mathrm{E} \\ \mathrm{E} \end{pmatrix}_\mathrm{cyl}. \label{eqn:aux_cross_r}
\end{equation}
where the final expression indicates the parity (even/odd) under the assumption of stellarator symmetry.  This vector $\mathbf{v}$ is parallel to the conventional Frenet binormal \citep{animov2001differential,mathews1964mathematical}, $\hat{\pmb \tau}_\text{F}=\mathbf{v}/|\mathbf{v}|$, and the curvature can be defined as,
\begin{equation}
    |\kappa|=\frac{1}{(\ell^\prime)^3}|\mathbf{v}|=\frac{1}{(\ell^\prime)^3}\sqrt{\left(v^R\right)^2+\left(v^\phi\right)^2+\left(v^z\right)^2},
\end{equation}
where we have written the components of $\mathbf{v}$ explicitly. Consider then a flattening point $\mathbf{r}_\mathrm{axis}(\phi_0)$ at which $\kappa=0$, which we take at a stellarator symmetric point. We characterize such points by a natural number $\nu\in\mathbb{N}_{>0}$ which we call the \textit{order} of the flattening point or of the zero of curvature. This is the order of the first non-vanishing $\phi$ derivative of the curvature $\kappa$,
\begin{equation}
    \nu=\min\left[ \{n\in\mathbb{N}^\mathrm{even} : \partial_\phi^n v^\phi \neq 0~\mathrm{or}~\partial_\phi^n v^z \neq 0\} \cup \{n\in\mathbb{N}^\mathrm{odd} : \partial_\phi^n v^R \neq 0\}\right],
\end{equation}
where parity of the various terms has been used here. This means that, in an alternating way, higher and higher derivatives of the components of $\mathbf{v}$ vanish locally up to an order. A natural way of formulating this requirement is by considering a local Taylor expansion, 
\begin{equation}
    v^j(\phi)=\sum_{n=0}^\infty v^j_{n}\frac{(\phi-\phi_0)^{n}}{n!}.
\end{equation}
In this form, the conditions for a zero of order $\nu=2k$ become the vanishing of all $\{v^\phi_{0},v^z_{0},v^R_{1},v^\phi_{2},v^z_{2},\dots,v^R_{2k-1}\}$ (and similarly for odd $\nu$). These coefficients can be directly related to the coefficients of the expansion of $R$ and $Z$ ({\em i.e.} $R_n$ and $Z_n$) through Eq.~(\ref{eqn:aux_cross_r}) using Cauchy products. Before explicitly doing so, let us note one important feature of $\mathbf{v}$, namely that
\begin{equation}
    Rv^\phi+Z'v^z=-R'v^R. \label{eqn:simp_v_rel}
\end{equation}
This implies an important relation between the triplet $\{v^\phi_{2k},v^z_{2k},v^R_{2k-1}\}$. When imposing condition $v^z_{2k}=0$, it is unnecessary to impose $v^\phi_{2k}=0$, because this is guaranteed by $v^R_{2k-1}=0$, which is already satisfied if we satisfy the constraints bottom-up. Thus, for a zero of order $\nu=2k$ all we need to satisfy is $\{v^z_{0},v^R_{1},v^z_{2},\dots,v^R_{2k-1}\}$ (and similarly for odd $\nu$). 
\par
The $v^z_{n}$ components can be explicitly computed, and the condition of them vanishing expressed as a constraint on $R_{n+2}$,
\begin{equation}
    R_{n+2}=R_n+\frac{1}{R_0}\sum_{k=0}^{n-1}\begin{pmatrix}
        n \\ k \end{pmatrix}[2R_{k+1}R_{n-k+1}-R_{n-k}(R_{k+2}-R_k)], \label{eqn:R_constr}
\end{equation}
for $R_1=0$. Thus, all higher order $R_{2k}$ can be reduced down to some multiple of $R_0$, where that factor grows exponentially. Proceeding similarly for $v^R$, we obtain a recursion for the components of $Z$,
\begin{equation}
    Z_{n+2}=-\frac{1}{R_0}\sum_{k=0}^{n-1}\begin{pmatrix}
        n \\ k \end{pmatrix}(R_{n-k}Z_{k+2}-2R_{n-k+1}Z_{k+1}), \label{eqn:Z_constr}
\end{equation}
in this case for $Z_0=0$. These conditions give the coefficients shown in \cite{camacho-mata_plunk_2022}, and can be readily computed numerically. In summary, for a zero of order $\nu$, we need to constraint locally coefficients up to ${R_{2\lceil \nu/2 \rceil},Z_{\lfloor \nu/2+1 \rfloor}}$ (excluding $R_0$ and $Z_1$). 
\par 
Depending on how $R$ and $Z$ are constructed, imposing these constraints becomes more or less involved. A straightforward representation that naturally captures the closure of the curve is a Fourier representation of both $R$ and $Z$,
\begin{equation}
    R(\phi)=\sum_{n=0}^{N_R}R_n^c\cos(nN\phi), \quad Z(\phi)=\sum_{n=0}^{N_Z}Z_n^s\sin(nN\phi). \label{eqn:RZ_fourier}
\end{equation}
To relate these coefficients to the local expansion ones, we simply need to use the Taylor expansions of the sine and cosine. Distinguishing between the local expansions about the mid-point $\phi_0=0$ and $\pi/N$, we would have,
\begin{subequations}
\begin{equation}
    R_{2k}[0]=(-1)^k\sum_{n=0}^{N_R}R_n^c(nN)^{2k}, \quad Z_{2k+1}[0]=(-1)^k\sum_{n=0}^{N_Z}Z_n^s(nN)^{2k+1},
\end{equation}
\begin{equation}
    R_{2k}[\pi/N]=(-1)^k\sum_{n=0}^{N_R}(-1)^nR_n^c(nN)^{2k}, \quad Z_{2k+1}[\pi/N]=(-1)^k\sum_{n=0}^{N_Z}(-1)^nZ_n^s(nN)^{2k+1}.
\end{equation}
\end{subequations}
Thus, the local constraints translate into constraints on the Fourier components of the curve, which one may impose in a multitude of ways.
\par
\subsection{Torsion}
A commonly used definition of torsion can be written
\begin{equation}
    \tau=\frac{\mathbf{r}_\mathrm{axis}'\times\mathbf{r}_\mathrm{axis}''\cdot \mathbf{r}_\mathrm{axis}'''}{|\mathbf{r}_\mathrm{axis}'\times\mathbf{r}_\mathrm{axis}''|^2}.
\end{equation}
Note that the denominator is precisely $|\mathbf{v}|$, Eq.~(\ref{eqn:aux_cross_r}). We have a potential divergence whenever the curvature of the axis vanishes. Let us see precisely what occurs by evaluating the torsion at the flattening point. In this case, we may write the numerator as,
\begin{align}
\tau_\mathrm{num}&=\mathbf{r}_\mathrm{axis}'\times\mathbf{r}_\mathrm{axis}''\cdot \mathbf{r}_\mathrm{axis}''' \nonumber\\
&=-\frac{v^R(v^z)'}{R}-\frac{v^z}{R'}\left[v^R\left(1+\left(\frac{R'}{R}\right)^2-\frac{R''}{R}\right)+v^z\left(\frac{R'Z'}{R^2}-\frac{Z''}{R}\right)-\frac{R'}{R}(v^R)'\right],
\end{align}
where we made explicit use of Eq.~(\ref{eqn:simp_v_rel}). Although this is not the most succinct form to write it, it is most convenient to discuss the behavior at our flattening point. To start with, the vanishing conditions on $v^R$ and $v^z$ obtained above guarantee that $\tau_\mathrm{num}$ vanishes to the same order as the denominator. This is just a matter of counting non-zero powers of $\phi-\phi_0$. 
\par
Thus, we are left with a leading non-zero expression that we may compute explicitly. Just to be complete, let us also rewrite the denominator in a similar form,
\begin{align}
    \tau_\mathrm{den}=&|\mathbf{r}_\mathrm{axis}'\times\mathbf{r}_\mathrm{axis}''|^2 \nonumber \\
    =& \left[1+\left(\frac{R'}{R}\right)^2\right](v^R)^2+2v^zv^RZ'\frac{R'}{R}+\left[1+\left(\frac{Z'}{R}\right)^2\right](v^z)^2.
\end{align}
With this, we write explicitly, at the flattening point
\begin{equation}
    \tau(\phi_0)=\begin{cases}
        \begin{aligned}
            &\frac{1}{R_0}\left(\frac{3}{2}\frac{v^R_{2k+1}}{v^Z_{2k}}+5\frac{Z_1}{R_0}\right)\left[1+\left(\frac{Z_1}{R_0}\right)^2\right]^{-1} & \text{for}~\nu=2k, \\
            & -\frac{3}{2R_0}\frac{v^z_{2(k+1)}}{v^R_{2k+1}} & \text{for}~\nu=2k+1. 
        \end{aligned}
    \end{cases} \label{eqn:torsion_local}
\end{equation}
The correctness of this expression can be checked against the few examples presented in \cite{camacho-mata_plunk_2022}. Of these expressions, the denominators $v^Z_{2k}$ and $v^R_{2k+1}$ in both the even and odd order zero cases are particularly important, as they could lead to a locally large value of the torsion. Although these quantities are non-zero by the assumption that the zero of curvature is of order $\nu+1$, they indicate a key property of this class of flattened curves that heightens the sensitivity of torsion to small changes in the curve parameters.
\begin{figure}
    \centering
    \includegraphics[width=0.9\linewidth]{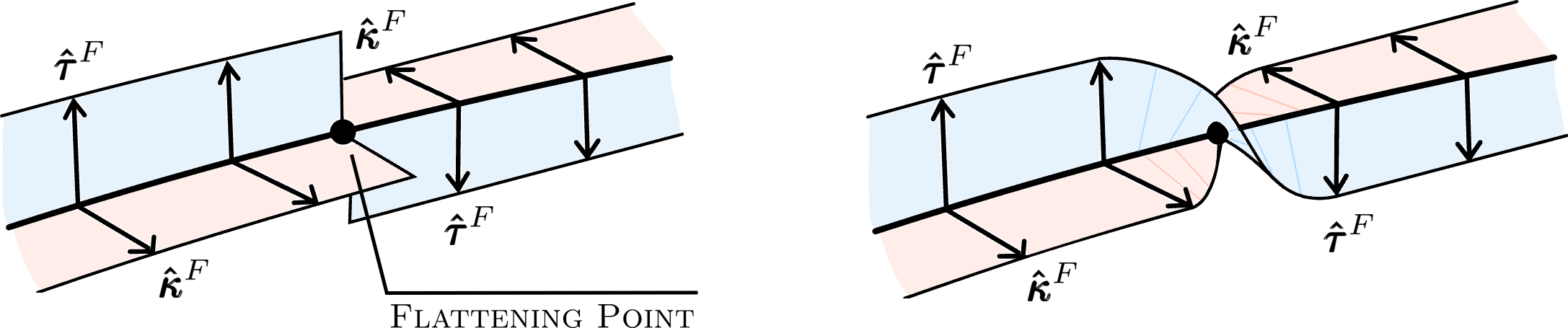}
    \caption{\textbf{Geometric explanation of diverging torsion.} The diagram represents the geometric mechanism behind the diverging torsion when the flatttening point is only approximately achieved. A discrete flip (left) of the frame in the ideal flattening point scenario becomes a continuous deformation (right) in a narrow region about the flattening point, and thus a large local value of torsion.}
    \label{fig:torsion_geo}
\end{figure}
\par
This sensitivity can be explained geometrically: Across a flattening point we know the Frenet-Serret frame normal to the curve to have a discrete sign flip when it is of odd order. With the introduction of a small error in this curve (for instance due to an imperfect numerical approximation), the discrete jump is replaced by a continuous rotation of the frame (see Figure~\ref{fig:torsion_geo}), occurring in a very narrow span, and therefore resulting in large torsion. 

Lets examine the related consequences of attempting to approximate a desired torsion and curvature of the axis by directly choosing functions $R(\phi)$ and $Z(\phi)$.  First, it is expected that any curve that is poorly represented with these coordinates ({\em e.g.} the figure-8) will exhibit large errors in torsion and curvature under truncation of its Fourier series representation, Eq.~(\ref{eqn:RZ_fourier}).  Even for cases that are, it can be observed from Equation \ref{eqn:torsion_local} that singularities in torsion can possibly arise whenever the axis curve lies close to the $z=0$ plane in the neighborhood of a flattening point; see the examples shown in Sec.~5.2 of \cite{camacho-mata_plunk_2022}. No such issues arise when using the Frenet-Serret construction as demonstrated with the low inclination curves in Section~\ref{sec:half-helicity-axes}.

\begin{figure}
    \centering
    \includegraphics[width=0.9\linewidth]{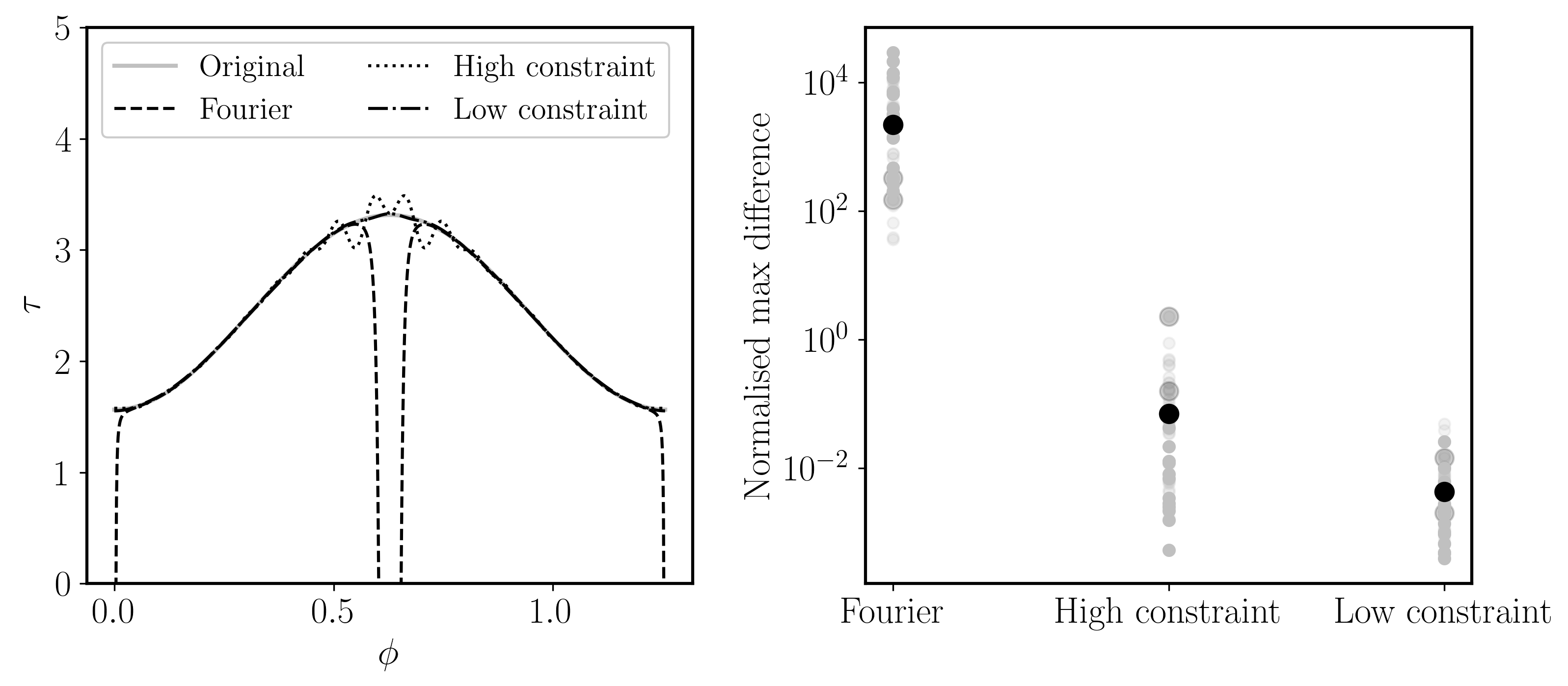}
    \caption{\textbf{Approximation to torsion of a curve.} In this figure we illustrate the attempts to reconstructing the torsion of a curve of the $(2,3)$ class using the Fourier cylindrical representation. The left plot shows a comparison of torsion of a few reconstruction attempts against the original value (grey). The reconstructions involve: (i) "Fourier" - simply Fourier decompose $R$ and $Z$ obtained from solving the Frenet-Serret equations, (ii) "High constraint" - the Fourier description of the curve enforcing the constraint on higher harmonics, and optimizing the remaining degrees of freedom, (iii) "Low constraint" - same as (ii) but using the lower harmonics to satisfy the constraints. The right plot quantifies the difference between the reconstructed and original torsion. The scatter represents the variability using different resolutions in the Fourier series as well as grid points. The black dot corresponds to the largest grid and Fourier resolution.}
    \label{fig:reconstructing_torsion}
\end{figure}

The problem is in fact more generic.  To illustrate this, lets consider a specific example not approaching these two limits.  Consider some curve of the class $(2,3)$ described in the Frenet form, closed and smooth. By solving the Frenet-Serret equations in Eqs.~(\ref{eqn:FS_eqns}), we may find the shape of the curve in $\mathbb{R}^3$, and thus we may obtain, numerically, functions $R$ and $Z$. Following the prescription above, we can represent $R$ and $Z$ as Fourier series, Eq.~(\ref{eqn:RZ_fourier}) to a desired level of accuracy. The presence of the flattening points means, however, that a numerical reconstruction can have very different torsion. Note how such reconstruction particularly struggles near the inflection points in Figure~\ref{fig:reconstructing_torsion}.
\par
In practice, we must perform such reconstruction imposing the constraints in Eqs.~(\ref{eqn:R_constr})-(\ref{eqn:Z_constr}) on the curve. Once care is taken of the neighborhood of the flattening point, the divergence of the torsion is avoided, and the reconstruction is much improved (see Figure~\ref{fig:reconstructing_torsion}). When it comes to constraining the coefficients describing the curve, and in order to avoid large harmonics, it is convenient to satisfy the constraint conditions through the lowest harmonics, and leave higher ones as degrees of freedom (the amplification of higher harmonics is otherwise apparent). After such optimization efforts, it is possible to capture the desired behavior of torsion.  The Frenet approach proposed in the main text of this paper avoids this complex overhead while giving complete control of both curvature and torsion, even away from flattening points.

\section{The first order approach and Mercier} \label{sec:app_merc}
It is the purpose of this section to shed some light on the alternative formulation of the $\sigma$-equation in which one uses elongation as an input to the problem rather than the magnetic field related $\bar{e}$.

Let us start by providing a geometric description of the "elongation profile" function $\rho$ defined in Eq.~(\ref{eq:rho-eqn})
\begin{equation}
    \rho = \bar{e} + \frac{1}{\bar{e}}(1 + \sigma^2), \tag{\ref{eq:rho-eqn}}
\end{equation}
which is monotonically related to the elongation $E$, Eq.~(\ref{eq:E-eqn}). Note that the validity of this expression goes beyond the special case of QI fields, so long as one appropriately adapts the definition of $\pelon$. Using standard near-axis notation,
\begin{equation}
    \pelon=\frac{B_{1c}^2+B_{1s}^2}{B_0\bar{B}}\frac{1}{\kappa^2}=\begin{cases}
    \begin{aligned}
        \frac{\eta^2}{\kappa^2}, & \quad \text{for QS} \\
        \bar{d}^2\frac{ B_0}{\bar{B}}, &\quad \text{for QI} 
    \end{aligned}
    \end{cases}
\end{equation}
following \cite{rodriguez2023magnetohydrodynamic}.
\par
\begin{figure}
    \centering
    \includegraphics[width=0.2\textwidth]{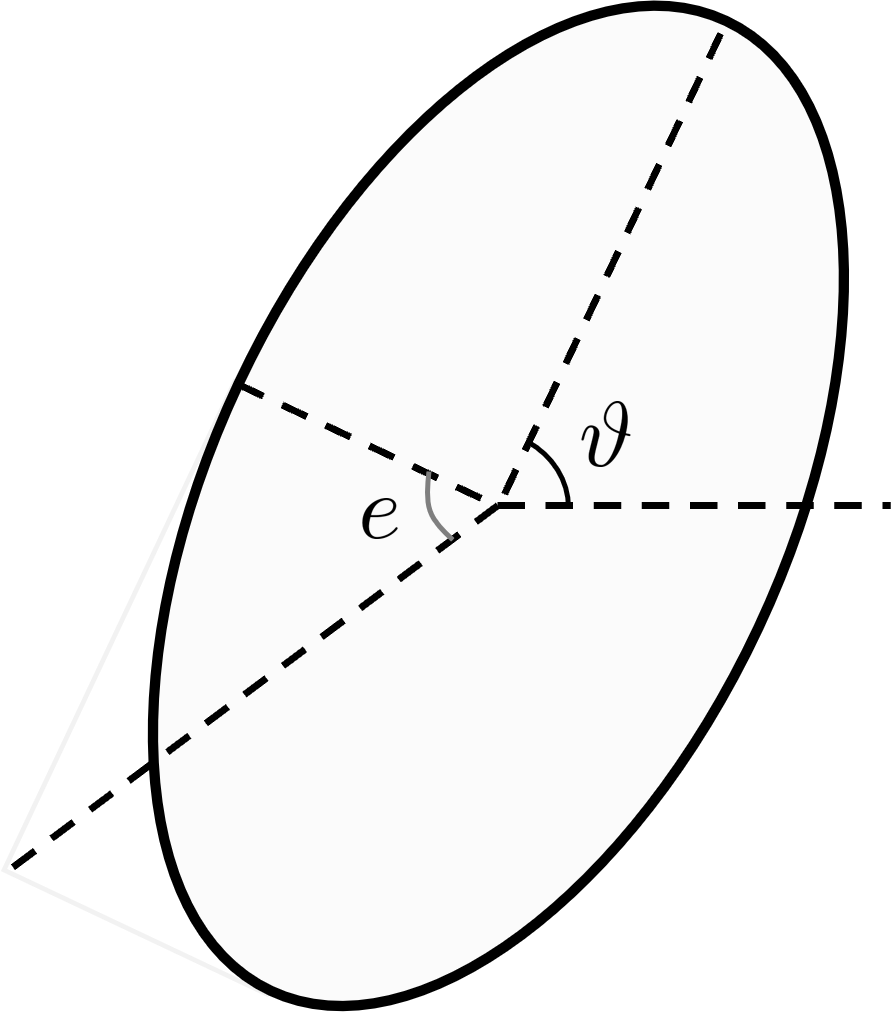}
    \caption{\textbf{Elliptical shapes and angles.} Diagram showing an ellipse framed in the normal Frenet-Serret frame where the ellipse rotation angle $\vartheta$ and elongation angle $e$ are defined. These two angles uniquely characterize ellipses (up to a scale). }
    \label{fig:ellipseAng}
\end{figure}
The relation of $\rho$ to the elongation of the elliptical cross-sections can be made more direct by introducing an angle $e$ such that the elongation of the ellipse (ratio of major to minor axes) $E=\tan e$, for $e\in[\pi/4,\pi/2)$. We give the geometric interpretation of $e$ in Figure~\ref{fig:ellipseAng}. It may be shown \citep{rodriguez2023magnetohydrodynamic} that $\rho=2/\sin 2e$, meaning that $\rho\in[2,\infty)$ and that for a circular cross-section, $e=\pi/4$ and $\rho=2$. In this defined domain, $\sin 2e> 0$ and $\cos 2e\leq 0$, which will be important later.

Let us then consider writing the $\sigma$-equation \ref{eq:sigma} in terms of this convenient choice of parameter $\rho$. We first rewrite it in full
\begin{equation}
    \frac{\mathrm{d}\sigma}{\mathrm{d}\varphi}+\left(\bar{\iota}_0-\frac{\mathrm{d}\widetilde{\alpha}}{\mathrm{d}\varphi}\right)\left(\pelon^2+1+\sigma^2\right)-2\pelon\frac{\mathrm{d}\ell}{\mathrm{d}\varphi}\left(\frac{I_2}{\bar{B}}-\tau\right)=0.
\end{equation}
defining $\widetilde{\alpha} = \alpha + M\varphi$ and $\bar{\iota}_0 = \iotaslash_0 - M$ to explicitly separate the secular and periodic parts of $\alpha$.  This can be rewritten (see Eqn.~\ref{eq:ebar-vs-rho}) as,
\begin{equation}
    \frac{\mathrm{d}\sigma}{\mathrm{d}\varphi}+2\rho\left(1-\sqrt{1-4(1+\sigma^2)/\rho^2}\right)\left[\frac{\rho}{2}\left(\bar{\iota}_0-\frac{\mathrm{d}\widetilde{\alpha}}{\mathrm{d}\varphi}\right)-\frac{\mathrm{d}\ell}{\mathrm{d}\varphi}\left(\frac{I_2}{\bar{B}}-\tau\right)\right]=0. \label{eqn:sigma_rho}
\end{equation} 
The presence of the square root has a rather trigonometric flavor. In fact, the whole formulation of the $\sigma$-equation involving elongation is evocative of the Mercier approach to the near-axis expansion. In the Mercier approach, the construction of the field involves not only the elongation of the surfaces, but also the rotation of the elliptical cross sections with respect to the Frenet frame, which is provided as input to the construction. Introducing this angle of rotation $\vartheta$ can transform Eq.~(\ref{eqn:sigma_rho}) even further as shown below.

\subsection{Mercier form of rotational transform}
Let us then try to recast the $\sigma$-equation in a form where the classic Mercier expression for the rotational transform of a field on axis is apparent. We introduce Mercier's $\eta_\mathrm{M}$, defined such that the elongation of the ellipses is given by $E=e^{\eta_\mathrm{M}}$. In the inverse coordinate near-axis, this means,
\begin{equation}
    \eta_\mathrm{M}=\ln\left(E\right).
\end{equation}
Now, knowing what the typical form of Mercier's $\iota_0$ expression is, we expect to find $\eta_\mathrm{M}$ involved in the problem through,
\begin{equation}
    \cosh\eta_\mathrm{M}=\frac{1}{2}\left(\tan e+\frac{1}{\tan e}\right)=\frac{1}{\sin 2e}=\frac{\rho}{2}.
\end{equation}
Thus, there is a simple relation between Mercier's elongation measure and the inverse-coordinate near-axis quantities. In fact, with this nice immediate relation to $e$, we may compute $\cos 2e=-\tanh\eta_\mathrm{M}$ and similarly (note the choice of sign). 
\par
With this in mind, and writing
\begin{align}
    \frac{1}{2\pelon}\frac{\mathrm{d}\sigma}{\mathrm{d}\varphi}&=\left(1-\frac{1}{\pelon\sin 2e}\right)\frac{\mathrm{d}\vartheta}{\mathrm{d}\varphi}-\frac{1}{\pelon}\frac{\sin2\vartheta}{\sin^2 2e}\frac{\mathrm{d}e}{\mathrm{d}\varphi}, \label{eqn:sigma_rot_eq_transformation}
\end{align}
the $\sigma$-equation so far reads,
\begin{equation}
    \frac{1-(1/\pelon)\cosh\eta_\mathrm{M}}{\cosh\eta_\mathrm{M}}\frac{\mathrm{d}\vartheta}{\mathrm{d}\varphi}-\frac{1}{\pelon}\frac{\sin2\vartheta}{\sin 2e}\frac{\mathrm{d}e}{\mathrm{d}\varphi}+\left(\bar{\iota}_0-\frac{\mathrm{d}\widetilde{\alpha}}{\mathrm{d}\varphi}\right)-\frac{\mathrm{d}\ell}{\mathrm{d}\varphi}\frac{1}{\cosh\eta_\mathrm{M}}\left(\frac{I_2}{\bar{B}}-\tau\right)=0,
\end{equation}
where we have divided through by $\cosh\eta_\mathrm{M}$. Knowing where we are headed, it is natural to separate the multiplying factor of $\mathrm{d}\vartheta/\mathrm{d}\varphi$ into a piece like $(1-\cosh\eta_\mathrm{M})/\cosh\eta_\mathrm{M}$ and the remainder. This remainder, together with $\mathrm{d}e/\mathrm{d}\varphi$, may then be expressed explicitly in terms of $\vartheta$ and $e$. The remaining term can be collected as an exact differential, such that the equation may be finally written as,
\begin{equation}
     \bar{\iota}_0+\frac{\mathrm{d}}{\mathrm{d}\varphi}\left[\vartheta-\arctan\left(\frac{\cos(e-\vartheta)}{\cos(e+\vartheta)}\right)-\widetilde{\alpha}\right]=-\frac{1-\cosh\eta_\mathrm{M}}{\cosh\eta_\mathrm{M}}\frac{\mathrm{d}\vartheta}{\mathrm{d}\varphi}+\frac{\mathrm{d}\ell}{\mathrm{d}\varphi}\frac{1}{\cosh\eta_\mathrm{M}}\left(\frac{I_2}{\bar{B}}-\tau\right).
\end{equation}
This is almost ready to be integrated from $\varphi=0$ to $2\pi$ to yield Mercier's expression for the rotational transform. However, we must be careful with the $\mathrm{d}/\mathrm{d}\varphi$ term in the left. If all terms inside the differential operator were periodic (like $\widetilde{\alpha}$ is), then we could simply drop the contribution to the total integral. However, there are clearly non-secular terms. After a total toroidal turn, $\vartheta$ can increase by $m\pi$, and thus $\int(\mathrm{d}\vartheta/\mathrm{d}\varphi)\mathrm{d}\varphi=m\pi$ and not zero. Is there a discrepancy with Mercier?
\par
The answer is, of course, no. The reason: the arctan function has \textit{precisely} the same secular behavior as $\vartheta$ does, which therefore cancel each other out. To see this, note that the argument of the cosine in the denominator of the argument of the arctan, $e+\vartheta$, changes, upon a full toroidal transit by $m\pi$. With $e\in[\pi/4,\pi/2)$ this means that the denominator must vanish $m$ times (at least). We say at least because the argument may be non-monotonic. When the denominator vanishes, it follows that the arctan jumps to the next Riemann sheet upon crossing of a branch cut across the point at infinity \citep[Ch.~4.4]{abramowitz1948handbook}. In total, it does so $m$ times. Therefore, upon integration over $\varphi$ from $0$ to $2\pi$ we have,
\begin{equation}
    \iotaslash_0-M=\frac{1}{2\pi}\int_0^{2\pi}\frac{\mathrm{d}\varphi}{\cosh\eta_\mathrm{M}}\left[\frac{\mathrm{d}\ell}{\mathrm{d}\varphi}\left(\frac{I_2}{\bar{B}}-\tau\right)+(\cosh\eta_\mathrm{M}-1)\frac{\mathrm{d}\vartheta}{\mathrm{d}\varphi}\right]\,.
\end{equation}
This is the Mercier form for the rotational transform on axis \citep[Eqs.~(30)-(31)]{mercier-near-axis}\citep[Eq.~(44)]{helander-review}. Note though that the rate of change of the rotation angle $\vartheta$ has in the definition here the opposite sign to that in the usual expression: measured from the normal to the axis here, from the axis to the normal in the standard form.

\subsection{Explicitly solving for the rotation angle}
We may take this Mercier perspective all the way to the end to formulate our $\sigma$-equation as a $\vartheta$-equation instead. Doing so might alleviate some of the questions about the choice of sign in the square root for $\bar{e}$. By using $\vartheta$, and allowing it to be a secular function, the trigonometric functions could take such changes of sign into account. 

With that in mind, let us reconsider the $\sigma$-equation defining for simplicity $S\equiv\sin2\vartheta$ and $C\equiv\cos2\vartheta$, so as to write
\begin{equation*}
    \frac{\sin4e/4}{1+C\cos 2e}\frac{\mathrm{d}S}{\mathrm{d}\varphi}-\frac{S}{1+C \cos 2e}\frac{\mathrm{d}e}{\mathrm{d}\varphi}+\left(\bar{\iota}_0-\frac{\mathrm{d}\widetilde{\alpha}}{\mathrm{d}\varphi}\right)-\frac{\mathrm{d}\ell}{\mathrm{d}\varphi}\sin 2e\left(\frac{I_2}{\bar{B}}-\tau\right)=0,
\end{equation*}
which we may rearrange as,
\begin{equation}
    \frac{\mathrm{d}S}{\mathrm{d}\varphi}-\frac{4S}{\sin 4e}\frac{\mathrm{d}e}{\mathrm{d}\varphi}+\frac{4}{\sin4e}\left(1+C\cos2e\right)\left[\bar{\iota}_0-\frac{\mathrm{d}\widetilde{\alpha}}{\mathrm{d}\varphi}-\frac{\mathrm{d}\ell}{\mathrm{d}\varphi}\sin 2e\left(\frac{I_2}{\bar{B}}-\tau\right)\right]=0,
\end{equation}
where the $1/\sin 4e$ factors exhibit singularities when the cross-sections become circles (i.e., $e=\pi/4$). Using the transformation to Mercier $\eta_\mathrm{M}$, the equation may be rewritten using $\sin4e=-2\sinh\eta_\mathrm{M}/\cosh^2\eta_\mathrm{M}$ and $\mathrm{d}\eta_\mathrm{M}/\mathrm{d}\varphi=2\cosh\eta_\mathrm{M}\,\mathrm{d}e/\mathrm{d}\varphi$,
\begin{equation}
    \frac{\mathrm{d}S}{\mathrm{d}\varphi}+S\underbrace{\frac{1}{\tanh\eta_\mathrm{M}}\frac{\mathrm{d}\eta_\mathrm{M}}{\mathrm{d}\varphi}}_{\equiv T(\varphi)}-2\left(\frac{1}{\tanh\eta_M}-C\right)\underbrace{\left[\left(\bar{\iota}_0-\frac{\mathrm{d}\widetilde{\alpha}}{\mathrm{d}\varphi}\right)\cosh\eta_\mathrm{M}-\frac{\mathrm{d}\ell}{\mathrm{d}\varphi}\left(\frac{I_2}{\bar{B}}-\tau\right)\right]}_{\equiv D(\varphi)/2}=0.
\end{equation}
Defining $\mathbb{S}(\varphi)=-D/\tanh\eta_\mathrm{M}$, the equation is
\begin{equation}
    \frac{\mathrm{d}}{\mathrm{d}\varphi}S+T(\varphi)S+D(\varphi)C+\mathbb{S}(\varphi)=0.\label{eq:S-eqn}
\end{equation}
This is the analogue to the $\sigma$-equation but in terms of the geometric elements of rotation and elongation. Practically, one may attempt to solve the equation for $\vartheta$ numerically. However, its very non-linear nature can make it numerically challenging to solve, especially when a solution with a secular $\vartheta$ is sought. Note that in this approach to the problem not every arrangement of elliptical cross-section around the magnetic axis is allowed as the elongation and ellipse rotation must satisfy this ODE (it can also be the case that multiple ones are.). This might appear contrary to the intuition from Mercier himself or more generally the direct near-axis expansion. However, by fixing $\alpha(\varphi)$ we are constraining the magnetic field and not all shapes will be consistent with that. More general forms of $\alpha(\varphi)$ can regain the full geometric freedom. 

For the present work we use to the $\sigma$-equation approach described in the main text, leaving the $\vartheta$ equation \ref{eq:S-eqn} for possible future exploration.

\bibliographystyle{jpp}
\bibliography{geom-qi}

\end{document}